\documentclass[a4paper,11pt]{article}
\usepackage{aaskaiid}

\newcommand{\Alfven}{$\rm Alfv\grave{e}n$}
\usepackage{orcidlink}

\usepackage{rotating}

\usepackage[utf8]{inputenc}
\title{Radio Wave Propagation as a Probe of the Solar Corona and Solar Wind}
\ShortTitle{Radio Wave Propagation as a Probe of the Solar Corona and Solar Wind}
\author[1]{K. Sasikumar Raja\,\orcidlink{0000-0002-1192-1804}}
\ShortName{Sasikumar Raja et al.} 
\author[2]{Prasad Subramanian\,\orcidlink{0000-0002-0459-231X}}
\author[3]{Susanta Kumar Bisoi\,\orcidlink{0000-0002-9448-1794}}
\author[4]{Janardhan Padmanabhan\,\orcidlink{0000-0003-2504-2576}}
\author[5]{Eduard Kontar\,\orcidlink{0000-0002-8078-0902}}
\author[6]{Anshu Kumari\,\orcidlink{0000-0001-5742-9033}}

\affiliation[1]{Indian Institute of Astrophysics, II Block, Koramangala, Bangalore-560 034, India}
\emailAdd{sasikumar.raja@iiap.res.in}

\affiliation[2]{Indian Institute of Science Education and Research, Pashan, Pune 411 008, India}
\emailAdd{p.subramanian@iiserpune.ac.in}

\affiliation[3]{Department of Physics and Astronomy, National Institute of Technology, 769008, Rourkela, India}
\emailAdd{bisois@nitrkl.ac.in}

\affiliation[4]{Physical Research Laboratory, Ahmedabad, India}
\emailAdd{jerry@prl.res.in}

\affiliation[5]{School of Physics and Astronomy, University of Glasgow, Glasgow G12 8QQ, UK}
\emailAdd{Eduard.Kontar@glasgow.ac.uk}

\affiliation[6]{Udaipur Solar Observatory, Physical Research Laboratory, Dewali, Badi Road, Udaipur - 313001, Rajasthan, India}
\emailAdd{anshu@prl.res.in}

\abstract{Radio waves propagating through an inhomogeneous, turbulent medium such as the solar corona and solar wind become distorted, causing the initially plane wavefronts becomes corrugated and acquire an RMS phase deviation across the wavefront. This leads to observable effects such as angular broadening of radio sources or intensity scintillation. Such waves can be used to probe the solar wind through various techniques, including angular broadening and interplanetary scintillation observations.
Such observations enable the study of several key properties, such as the phase structure function, amplitude of turbulence, density modulation index, solar wind heating rates, magnetic field topology, and dissipation scales. These phenomena provide critical insights into the physical processes governing the solar corona and solar wind and its interaction with radio waves, offering valuable constraints on both coronal and solar wind turbulence and coronal magnetic field configurations.
Currently, the limited number of radio sources near the ecliptic restricts our observations. However, the SKA-Low and SKA-Mid are expected to detect a significantly larger number of radio sources, thereby providing deeper insights into the solar corona, solar wind, and heliosphere. Long-term observations will be crucial to understanding how the above-mentioned parameters vary with heliocentric distance and over the solar cycle.
}	
	
\begin{document}
\maketitle

\section{Introduction}

The Square Kilometre Array Observatory (SKAO) consists of the SKA-Low, which operates from 50–350 MHz, and the SKA-Mid, which operates from 0.35–15.4 GHz. Both arrays provide extensive short-baseline coverage, which is crucial for studying solar wind density turbulence including parameters such as the amplitude of turbulence, density fluctuations, dissipation scales, and proton heating rates. At present, only a few radio sources are suitable for radio occultation observations due to limited sensitivity. Once the SKAO is commissioned, many more such sources will become available, enabling detailed probing of the solar corona and solar wind. Note that the expected continuum sensitivity of SKA is at the $\mu$Jy level for typical integrations and reaches sub-$\mu$Jy levels for deep observations. For example, SKA-Low achieves $\sim$2--5~$\mu$Jy~beam$^{-1}$ in 1~hr, while SKA-Mid reaches $\sim$1--2~$\mu$Jy~beam$^{-1}$, improving to $\sim$0.05--0.1~$\mu$Jy~beam$^{-1}$ in deep integrations \citep{Dewdney2013,Braun2019}. This article presents various science cases that can be explored with the SKAO.
 
Despite several decades of intense research through observations and modelling, solar wind heating, origin and evolution, and acceleration of solar wind is still poorly-understood. There are various techniques available to probe the solar wind. Most of the solar wind theories treat incompressible turbulence where as solar wind exhibits density fluctuations that impact the plane waves as refractive index changes. Understanding turbulence helps interpreting variety of observations and theories. It helps understanding and testing of compressibility of a solar wind [e.g. Tu et al 1994, Hnat et al 2005]. Radio sources when observed through the solar corona and solar wind broadens and that leads to the under estimation of the brightness temperatures. Therefore it leads to misinterpretation of the source properties and emission mechanism.  It helps understanding the dissipation mechanism in the solar wind and therefore the solar wind heating and acceleration of the solar wind. Turbulence plays a role in the propagation of Solar Energetic Particles (SEP) through out the heliosphere (e.g., Reid \& Kontar 2010).

In the solar wind, turbulent density inhomogeneities play a vital role in scattering of the radio waves \citep{Col1989, Yam98, Bis14, Mug2017, Kru2018, Sas2019b, Kru2020, Chrysaphi2026}.  The spectrum of turbulent density fluctuations ($P_{\delta n}$) in the solar wind is commonly modeled as a power law with an exponential falloff at the inner or dissipation scale \citep{Bas94, Bas95, Arm1995, Ale12, Ing2015, Sas2016, Sas2017}. When we consider anisotropy in the scatter-broadening images by allowing for different wavenumbers along ($k_{x}$) and perpendicular ($k_{y}$) to the large scale magnetic field \citep{Ing2015},

\begin{eqnarray}
P_{\delta n}(k, R) = C_{N}^{2}(R) (\rho^2 ~k_x^2+k_y^2)^{-\alpha/2} \\ \nonumber
 \times ~ exp\bigg\{-(\rho^2 ~k_x^2+k_y^2)\bigg({l_{i}(R) \over 2 \pi }\bigg)^{2}\bigg\} \,  ,
\label{eq1}
\end{eqnarray}

where $R$ is the heliocentric distance, $C_{N}^{2}$ represents the amplitude of the density spectrum and $\rho \equiv k_{x}/k_{y}$ is interpreted as the ratio of the major and minor axes in the angular broadened images. The inner scale is also assumed to be anisotropic, with $l_{ix} = \rho l_{iy}$ and $l_{i} \equiv \sqrt{l_{ix}^{2} + l_{iy}^{2}}$. The slope ($\alpha$) of the turbulent power spectrum is well known to follow the Kolmogorov scaling law ($\alpha = 11/3$) at large scales. Also it is known to flatten near the dissipation scale at $\alpha = 3$ before it turns over steeply at the inner scale \citep{Bas94, Col1989, Sas2016, Sas2017, Mug2017}. Since our observations often sample scales in the vicinity of dissipation scale, $\alpha = 3$ is used.

However, when the medium is isotropic, the turbulent spatial power spectrum ($P_{\delta N}$) is defined as \citep{Bas94, Ing2015}, 

\begin{equation}\label{eq:ss}
P_{\delta N}(k, R) = C_{N}^{2}(R) k^{-\alpha} \times \exp{-(kl_{i}(R) / 2 \pi)}^{2},
\end{equation}

where $k$ is the wavenumber, and $l_i$ is the inner/ dissipation scale.

It is worth mentioning that the injected large-scale energy in the solar wind breaks up into smaller scales until it is dissipated by heating the protons using gyro-resonant interactions. Also, note that the scales at which the energy is injected are called `outer scales', and the scales at which the dissipation happens are called `inner scales' \citep{Kul2005}. Using remote sensing observations, it is found that, at large scales, the density spectrum follows the Kolmogorov scaling law with $\alpha=11/3$ \citep{Col1989, Spa02}. However, at small scales, the spectrum flattens to $\alpha=3$ \citep{Col1989}. Further, it is important to note here that $C_N^2$ can be measured for both proton inertial scale model  \citep{Col1989,Lea1999,Lea2000,Smi2001,Che2014,Bru2014,Sas2019a} and proton gyroradius model \citep{Bal2005, Sah13, Bis14, Che14, Sas2019a}. 

\section{Observations}

In the radio occultation technique, a radio point source (say the Crab Nebula) can be observed through the foreground solar wind 
during June of each year, leading to the following observational characteristics:  
(i) the apparent angular size of the radio source increases due to scattering by the turbulent medium,  
(ii) since the intrinsic flux density of the source remains constant over long timescales, the peak flux density decreases 
as the apparent source size increases, while the integrated flux density remains constant,
(iii) for heliocentric distances below $\sim 10~R_{\odot}$, the radio source exhibits anisotropic broadening, 
enabling the measurement of the anisotropy parameter (i.e., the ratio of the major to minor axis of the source) 
\citep{Ble1972, Den1972, Sas2017}, and  
(iv) the position angle of the major axis of the source, measured from north through east, can also be determined.  

\begin{figure}[h]
    \centering
	\includegraphics[width=0.6\columnwidth]{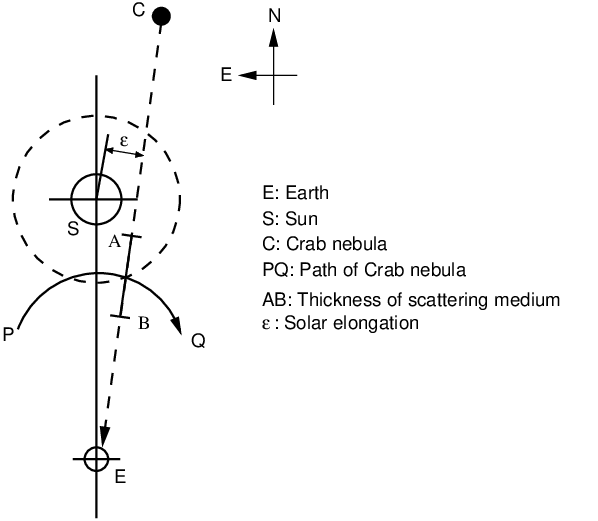}
    \caption{The schematic diagram shows the geometry of Crab nebula occultation; `PQ' indicates the projected path of the Crab nebula during the month of June every year. The closest point of `PQ' to `S' is $\approx$ $5~R_{\odot}$. The radiation from `C' passes through the effective turbulent medium `AB' at a solar elongation of `$\epsilon$' as viewed from `E'.}
    \label{fig:schematic}
\end{figure}

Since the Crab occultation technique is well established, we summarize here only the aspects relevant to this article. The Crab Nebula is typically observed using a single-element interferometer as it passes through the solar wind at heliocentric distances of approximately 10–45 $R_{\odot}$ during mid-June each year. As the source approaches the Sun, enhanced scattering caused by turbulent density irregularities in the solar wind leads to an increase in its apparent angular size. Eventually, the nebula becomes sufficiently broadened that it is resolved out by the interferometer, resulting in a drop of the measured visibility to undetectable levels which marks the so-called ``occultation".

\begin{figure}[!ht]
\centering
\includegraphics[width=14cm]{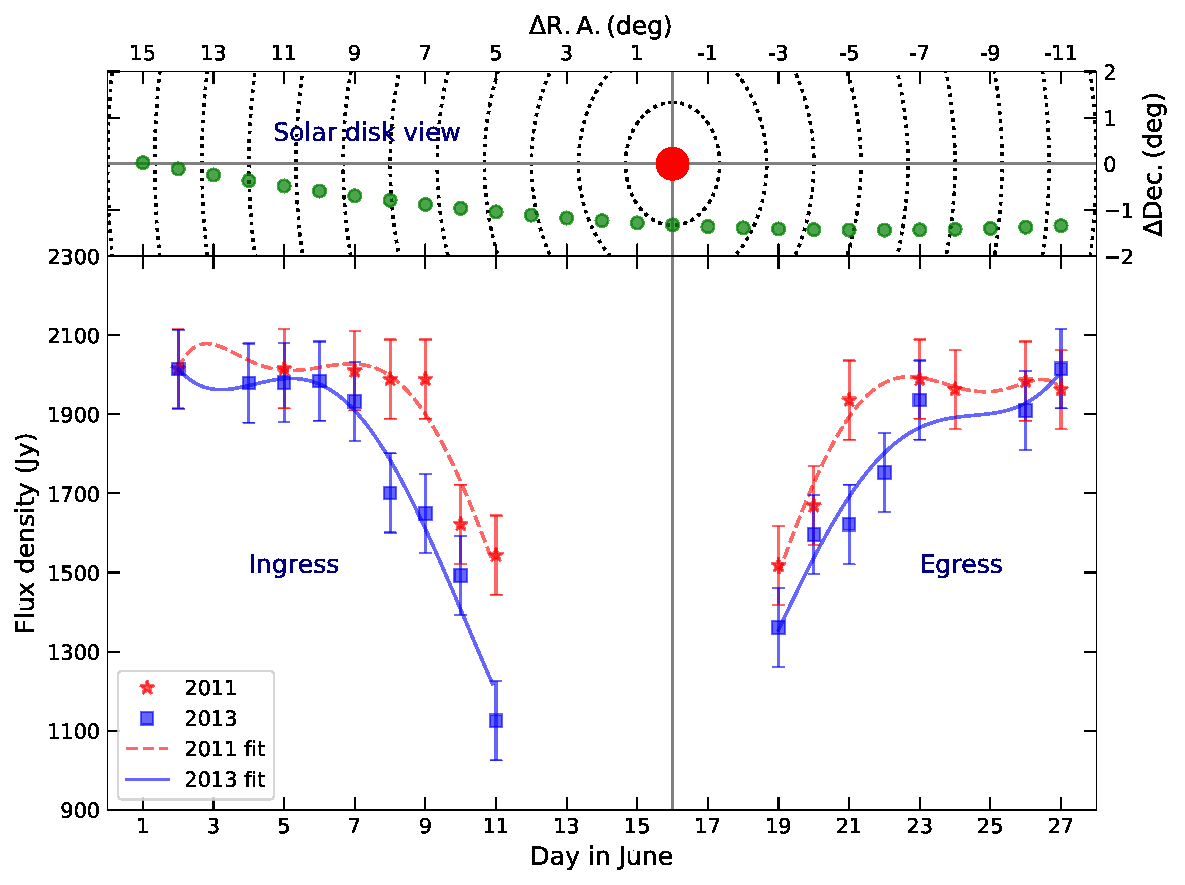}
\caption{The upper panel shows the solar disk view of the Crab nebula occultation technique. 
The filled red colored circle represents the Sun, while the green circles indicate the positions of the Crab nebula relative to the Sun on different dates. 
Here, $\Delta$R.A. and $\Delta$Dec denote the offset distances of the Crab nebula from the Sun in right ascension and declination, respectively. 
The innermost concentric circle around the Sun has a radius of $5~R_{\odot}$, and the radii of the subsequent circles increase by $5~R_{\odot}$ each. 
The lower panel presents the observed flux densities of the Crab nebula on various days during its occultation by the foreground solar corona. 
The periods before and after June 16 correspond to the ingress and egress phases, respectively. 
The data points marked with `*` and `square' correspond to the measurements from June 2011 and June 2013, respectively.
}
\label{fig:figure2}
\end{figure}

Figure \ref{fig:schematic} is a schematic diagram showing the occulting geometry during observations of the Crab Nebula. The lower panel of Figure \ref{fig:figure2} presents the variation of the observed flux density of the Crab Nebula during June 2011 and 2013, while the upper panel illustrates the corresponding solar disk geometry of the occultation. In 2011, the observed flux density shows a steady decline from its pre-occultation level of approximately $2015 \pm 100$ Jy beginning on 10 June (at $ \approx 23 R_{\odot}$) during ingress. In contrast, the decrease begins earlier, on 8 June (at $ \approx 30 R_{\odot}$), in 2013 (see Figure \ref{fig:figure2}). A similar trend is observed during the egress phase. The pre-occultation flux is regained around 21 June ($ \approx 21 R_{\odot}$) in 2011 and around 23 June ($ \approx 29 R_{\odot}$) in 2013. No interference fringes were detected between 12 and 18 June in either year, corresponding to heliocentric distances of about $ \approx 15 R_{\odot}$ (ingress, 12 June) and $ \approx 10 R_{\odot}$ (egress, 18 June).

Although the heliographic latitudes sampled by the line of sight to the Crab Nebula differ between ingress and egress \citep{Kun65}, the occultation profiles for 2011 and 2013 (Figure \ref{fig:figure2}) appear fairly symmetric. This behavior is expected, as the solar cycle 24 maximum occurred in 2013, when the distribution of solar wind density fluctuations is known to be nearly spherically symmetric \citep{Man93}.

In this study, we utilize these observations along with similar data obtained in previous years. The earliest Crab Nebula occultation measurements were reported by \citet{Mac1952} in 1952 at 38 and 80.5 MHz. Subsequent observations between 1952 and 1958 were made at 38, 81, and 158 MHz using baselines ranging from 60 to 1000 m \citep{Hew57, Hew58}. Additional measurements at 26.3 and 38 MHz, over baselines of approximately 700–1630 m, were reported for 1961–1962 by \citet{Hew63}. The normalized visibilities from these earlier data, obtained across different frequencies and baselines, have been scaled to 80 MHz and a 1.6 km baseline using the general structure function described in \S \ref{sf1}.

\section{Structure function and amplitude of turbulence ($C_N^2$)}

The building block for obtaining radio images is the quantity $\Gamma(s)$ that measures the spatial coherence of the electric field detected by a pair of antennas separated by a distance $s$: 

\begin{equation}\label{mf}
\Gamma(s)={V(s) \over V(0)} = {\langle E(r)E^*(r+s) \rangle \over \langle |E|^2 \rangle}
\end{equation}
The quantity $V(s)$, called the visibility, is the time-averaged correlation between electric fields detected by a pair of antennas separated by a distance $s$ and $V(0)$ is the so-called ``zero-spacing'' visibility. Each pair of antennas in an interferometric array yields a visibility, which is a point in the Fourier transform of the scatter-broadened image of the object.

The structure function, $D_{\phi}(s)$, characterizes the phase perturbations introduced by density inhomogeneities in the medium and is expressed as \citep{Pro75, Ish78, Col89, Arm90}
\begin{equation}
\Gamma(s) = e^{-D_{\phi}(s)/2} \, .
\label{eq4}
\end{equation}
Alternatively,
\begin{equation}
D_{\phi}(s) = -2 \ln \Gamma(s) = -2 \ln \left[ \frac{V(s)}{V(0)} \right] \, ,
\label{eq3}
\end{equation}
where $V(s)$ and $V(0)$ denote the ensemble-averaged visibilities. Note that $V(0)$ corresponds to the flux density of the Crab nebula when it is far from the Sun. Crab occultation observations are typically carried out using a single baseline with a specific value of $s$.  

\subsection{The General Structure Function (GSF)}\label{sf1}

Over the years, theoretical developments and observations have converged on a widely accepted formulation for the structure function that describes density fluctuations in the solar corona and solar wind \citep[e.g.,][]{Col87, Arm00, Bas94, Pra11}.  
These formulations, however, are valid only in the asymptotic limits where the observing baseline $s$ is either much smaller or much larger than the inner scale $l_i(R)$. Such approximations may not hold in some cases, since depending on the assumed model for the inner scale, the observing baseline $s$ can be comparable to $l_i(R)$. Under these circumstances, the use of asymptotic expressions leads to inaccuracies, necessitating the use of the \textit{General Structure Function} (GSF), which remains valid for all regimes: $s \ll l_i(R)$, $s \approx l_i(R)$, and $s \gg l_i(R)$ \citep{Ing2015}.  

Scatter-broadened images of radio sources viewed through the solar wind are observed to be anisotropic only at heliocentric distances $\leq 5$--6~$R_{\odot}$ \citep{Ana1994, Arm90}. However for observations that were obtained over the range 10--45~$R_{\odot}$, it is sufficient to employ the isotropic form of the GSF, defined as
\begin{eqnarray}
\label{gsf}
D_{\phi}(s) &=& \frac{8 \pi^2 r_e^2 \lambda^2 \Delta L}{2^{\alpha-2}(\alpha-2)} 
\Gamma \left( 1 - \frac{\alpha-2}{2} \right)
\frac{C_N^2 (R) \, l_i(R)^{\alpha-2}}{1 - f_p^2 (R) / f^2} \\ \nonumber
&& \times \left\{ \, {_1F_1} \!\left[ -\frac{\alpha-2}{2},\, 1,\, -\left( \frac{s}{l_i(R)} \right)^2 \right] - 1 \right\} \, ,
\end{eqnarray}
where ${_1F_1}$ is the confluent hypergeometric function, $r_e$ is the classical electron radius, $\lambda$ is the observing wavelength, $R$ is the heliocentric distance, $\Delta L$ is the thickness of the scattering medium, and $f_p$ and $f$ denote the plasma and observing frequencies, respectively.  

The functional form of the structure function is therefore well established. The visibilities from Crab occultation observations provide a constraint on its amplitude. Since the structure function depends explicitly on both observing wavelength and baseline, such dependence was used to normalize the visibilities obtained at different observing frequencies and baselines to a common reference of 80~MHz and a 1.6~km baseline.  

The origin of the inner (dissipation) scale, $l_i(R)$, remains an active topic of research.  
Some studies identify it with the proton inertial length \citep{Col89, Har89, Yam98, Ver96, Lea99, Lea00, Smi01, Bru14}, while others associate it with the proton gyroradius \citep{Bal2005, Sah13}.  
  
In several cases, the observing baseline is comparable to the inner scale.  
Note that the baseline used in the 2011 and 2013 observations ($s = 1600$~m) is comparable to the proton gyroradius within the heliocentric distance range $\approx$10--45~$R_{\odot}$.  
However, when the proton inertial length is adopted for the inner scale, the corresponding baseline lengths are significantly smaller.  
Therefore, GSF can be employed that remains valid across all these regimes: from $s \ll l_i(R)$ through $s \approx l_i(R)$ and up to $s \gg l_i(R)$.

\subsection{Estimating the scattering measure}

The scattering measure (SM) is defined as the path integral
\begin{equation}
\label{smdef}
{\rm SM} = \int C_{N}^{2}(R) \, dl \approx C_{N}^{2}(R) \, \Delta L\, ,
\end{equation}
where the integration is carried out over the depth over which scattering takes place. When the scattering is confined to a thin screen,
the approximation indicated in Eq~\ref{smdef} is acceptable, where $\Delta L$ is the thickness of the scattering screen.
We use the GSF to calculate the scattering measure, which in turn will be used to determine $C_{N}^{2}(R)$. 

\begin{eqnarray}\label{eq:sm1}
{\rm SM} &=& C_{N}^2(R) \Delta L \\ %
	 &=& \left(f(\alpha,\lambda) \, \frac{l_i(R)^{\alpha-2}}{r_f(R,\lambda)} \, \left(_1F_1(\alpha,s,R)-1\right) \right)^{-1} D_{\phi}(s) \nonumber
\end{eqnarray}

where,
\begin{eqnarray*}
f(\alpha, \lambda) &=& \frac{8\pi^2r_e^2\lambda^2}{2^{\alpha-2}(\alpha-2)} \Gamma\left(1-\frac{\alpha-2}{2}\right), \\
r_f(R,\lambda) &=& 1-f_p^2(R)/f^2, \\
_1F_1(\alpha,s,R) &=& _1F_1\left[-\frac{\alpha-2}{2},1,-\left(\frac{s}{l_i(R)}\right)^2 \right],
\end{eqnarray*}

\subsection{Heliocentric dependence of $C_{N}^{2}$}
The structure function can be used to calculate the scattering measure (Eq~\ref{eq:sm1}). The SM using turbulence amplitude $C_N^2(R)$ at different solar elongations $R_{0}$ is described below.  Assuming solar wind turbulence at these heliocentric distances ($10-45~R_{\odot}$) to be spherically  symmetric, the SM can also be written as \citep{Spa95}:

\begin{equation}
{\rm SM} = \int_0^\infty \,\, C_N^2(R)\,\, {\rm d}R = \frac{\pi}{2}C_{N}^2(R_{0})\, R_{0}
\end{equation}

\begin{equation}\label{eq:sm2}
C_{N}^2(R_{0}) = \frac{2}{\pi}\frac{\rm SM}{R_{0}}
\end{equation}

 where $C_{N}^2(R_{0})$ denotes the amplitude of density turbulence at impact parameter $R_0$. The impact parameter $R_0$ is related to 
 the solar elongation in solar radius in arc minute. Comparing with 
 Eq~\ref{smdef} shows that the scattering screen thickness is found as
$\Delta L = (\pi/2) R_0$, in computing $C_{N}^2(R_{0})$ from the SM.

The SM is estimated from the observed structure function ($D_{\phi}(s)$) using two inner scale models: the proton inertial length and and proton gyroradius. Furthermore, 
the SM depends upon the assumed value of power  
law index ($\alpha$) of the density fluctuation spectrum. Generally, the spectrum is observed to follow a Kolmogorov-like scaling with $\alpha = 11/3$.
However, there is also some evidence for local flattening of the density fluctuation spectrum at large wave numbers \citep{Cel87,Col89, Bas94};
some authors therefore use $\alpha = 3$. 

\begin{figure}[!ht]
\centering
\includegraphics[width=12cm]{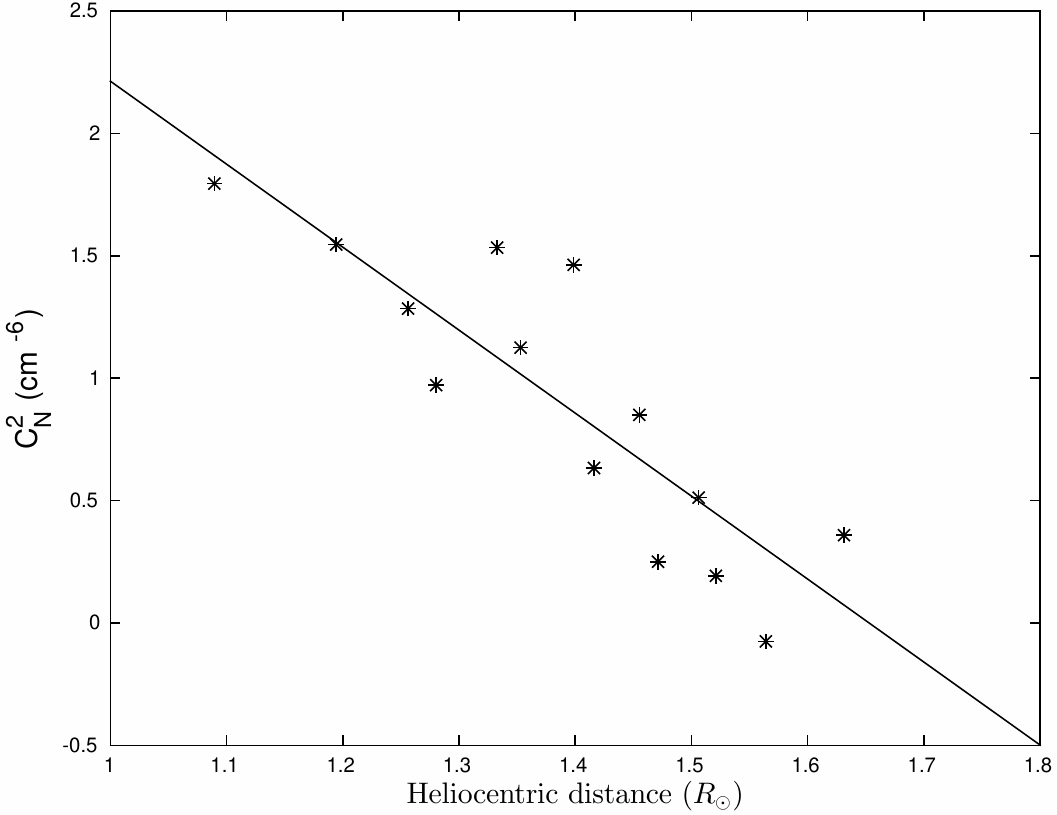}
\caption{A log-log plot of $C_{N}^{2}$ vs the heliocentric distance (in $R_{\odot}$) derived from observations in 2013. We used $\alpha=3$ and
the inner scale is the proton inertial length. The fit to $C_N^2(R)=A~R^{-\gamma}$ yields $\gamma = 3.4$ and $A = 4 \times 10^{5}~{\rm cm}^{-6}$.}
\label{fig:cn21}
\end{figure}

Using all the existing data in the literature, $C_{N}^2$ was computed as a function of heliocentric distance between 10 and 45 $R_{\odot}$. Since the observation span years corresponding to solar minimum as well as solar maximum, the data from each year was studied separately. For instance, Figure \ref{fig:cn21} shows the variation of $C_N^2$ with heliocentric distance using data from 2013. A fit of a function of the form $C_N^2(R)=A~R^{-\gamma}$ is plotted in these Figures. We find that the data in Figure \ref{fig:cn21} suggests $A = 4\times 10^5~{\rm cm}^{-6}$ and $\gamma = -3.4$ with a goodness of fit (adjusted $R^2$) 0.72. Such measurements were carried out for different years. For instance, in 2011, $C_{N}^{2}(R) = 3.2 \times 10^{4} R^{-2.8}$ for $\alpha = 3$ and the proton inertial length as the inner scale. On the other hand, $C_{N}^{2}(R) = 400 R^{-2.1}$ for $\alpha = 3$ and the proton gyroradius (with proton temperature = $10^5$ K) as the inner scale. To the best of our knowledge, the only such measurements in the literature so far is due to \citet{Spa95} and \citet{Spa96}, who determined the heliocentric dependence of $C_{N}^{2}$ from 10 to 60 $R_{\odot}$ using Very Long Baseline Interferometer (VLBI) observations during July and August 1991, which is $\approx$ 2 years past the maximum of cycle 22 in the declining phase. Their result, which assumes a Kolmogorov spectrum ($\alpha = 11/3$) is $C_{N}^{2}(R) = 3.81 R^{-3.66}$ in ${\rm cm}^{-20/3}$. Of the results we have compiled, data from 1960
corresponds to a similar phase in cycle 19. For this epoch, we obtain $C_{N}^{2} \propto R^{-\gamma}$, with $\gamma$ ranging from 3.2 to 3.3. These results 
thus yield a remarkably similar dependence of $C_{N}^{2}$ with heliocentric distance for the only instance in the literature where such a comparison can be made.

\subsection{Solar cycle dependence of $C_{N}^{2}(R)$}

\begin{figure*}
\centering
\includegraphics[width=12cm]{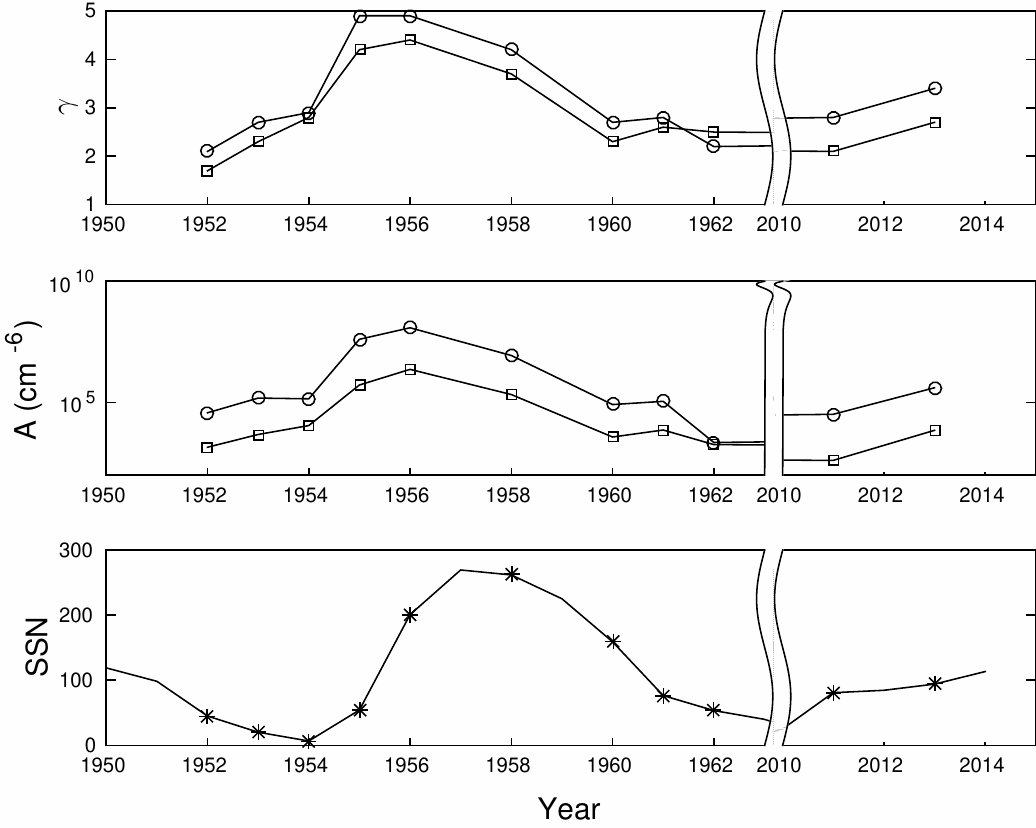}
\caption{The top panel and middle panel show $\gamma$ and $A~(cm^{-6})$ as a function of time. The `circles' and `squares' indicate the proton inertial and proton gyroradius inner scale models respectively with $\alpha=3$. For the proton gyroradius model, a temperature of $10^5$ K was used. 
The solid line in the bottom panel shows the yearly averaged sunspot number and the `*' represents 
the year in which the Crab occultation measurements were made. It is clearly seen that both $\gamma$ and $A$ correlates with the solar cycle.}
\label{fig:solarcycle_3}
\end{figure*}

The values of $A$ and $\gamma$ are significantly different for different observation years, which correspond to different phases of the solar cycle.
The solar cycle dependence of $A$ and $\gamma$ is shown in Figures \ref{fig:solarcycle_3}. 
The top and middle panels in the Figures \ref{fig:solarcycle_3} shows the temporal variation of $\gamma$ and $A$. 
For the comparison, the yearly averaged sunspot number (SSN)\footnote{http://www.sidc.be/silso/datafiles} for different years are plotted in the bottom panel 
of Figures \ref{fig:solarcycle_3}. Upon comparing the top and middle panels with the bottom ones, it is evident that both $A$ and $\gamma$ are well correlated with the sunspot number. These trends hold irrespective of whether the proton gyroradius model is used or proton inertial length model for the inner scale with $\alpha = 3$. 

The correlation between $A$ and the sunspot number is indicative of the fact that the overall magnitude of scattering is higher 
during solar maximum as compared to solar minimum. This is consistent with earlier results using interplanetary scattering observations 
\citep{Jan11, Man12, Jan15}. 

The correlation between $\gamma$ and the sunspot number indicates that the scattering strength falls off faster with heliocentric 
distance when solar activity increases. This might be because the large-scale solar magnetic field becomes more multipolar with increasing solar 
activity. For instance, this is reflected by the increasing complexity of the streamer belt with solar activity \citep{Wan00,Ric08}. 
Higher order multipolar fields are known to fall off more rapidly with heliocentric distance than a dipole, and this could be reflected in the spatial behavior 
of the scattering strength, characterized by $\gamma$. Conversely, it has been reported earlier \citep{Tok00} 
that the scintillation index for IPS observations shows a rather shallow variation with heliocentric distance towards solar minimum. 
It should also be borne in mind that the Crab nebula passes from low latitudes to high(er) ones (upper panel of Figure \ref{fig:figure2}). Near solar minimum, 
this means that it progresses from sampling the slow solar wind to the fast solar wind, and this is an additional complicating factor. 
Near solar maximum, the solar wind is relatively more symmetric with latitude, and is predominantly slow \citep{Mcc00, Asa98}.
%

\section{Density fluctuation index ($\epsilon_{N_e}=\delta N_{k_i} / N_e$)}\label{lab:densmod}
The density fluctuations $\delta N_{k_i}$ at the inner scale and spatial power spectrum are related as follows \citep{Cha2009}

\begin{equation}\label{eq:deltn}
{\delta}N_{k_i}^2(R) \sim 4 \pi k_i^3 P_{\delta N} (R, k_i) = 4 \pi C_{N}^{2}(R) k_i^{3 - \alpha} e^{-1} \,,
\end{equation}

where $k_{i} \equiv 2 \pi/l_{i}$. 

By knowing the ${\delta}N_{k_i}$ and the background electron density 
($N_{e}$), the density modulation index ($\epsilon_{N_e}$) can be measured using, 

\begin{equation}\label{eq:df}
\epsilon_{N_e}(R) \equiv {~\delta N_{k_{i}}(R) \over N_{e}(R)}. 
\end{equation}
 
For the sake of completeness, the measured density modulation indices and its variation with heliocentric distance is shown in  \citet{Sas2016}. Similarly, solar cycle dependence of the density modulation indices is shown in upper panel of Figure \ref{fig:sc} \citep{Sas2016}. 

\begin{sidewaysfigure}
	\includegraphics[width=23cm]{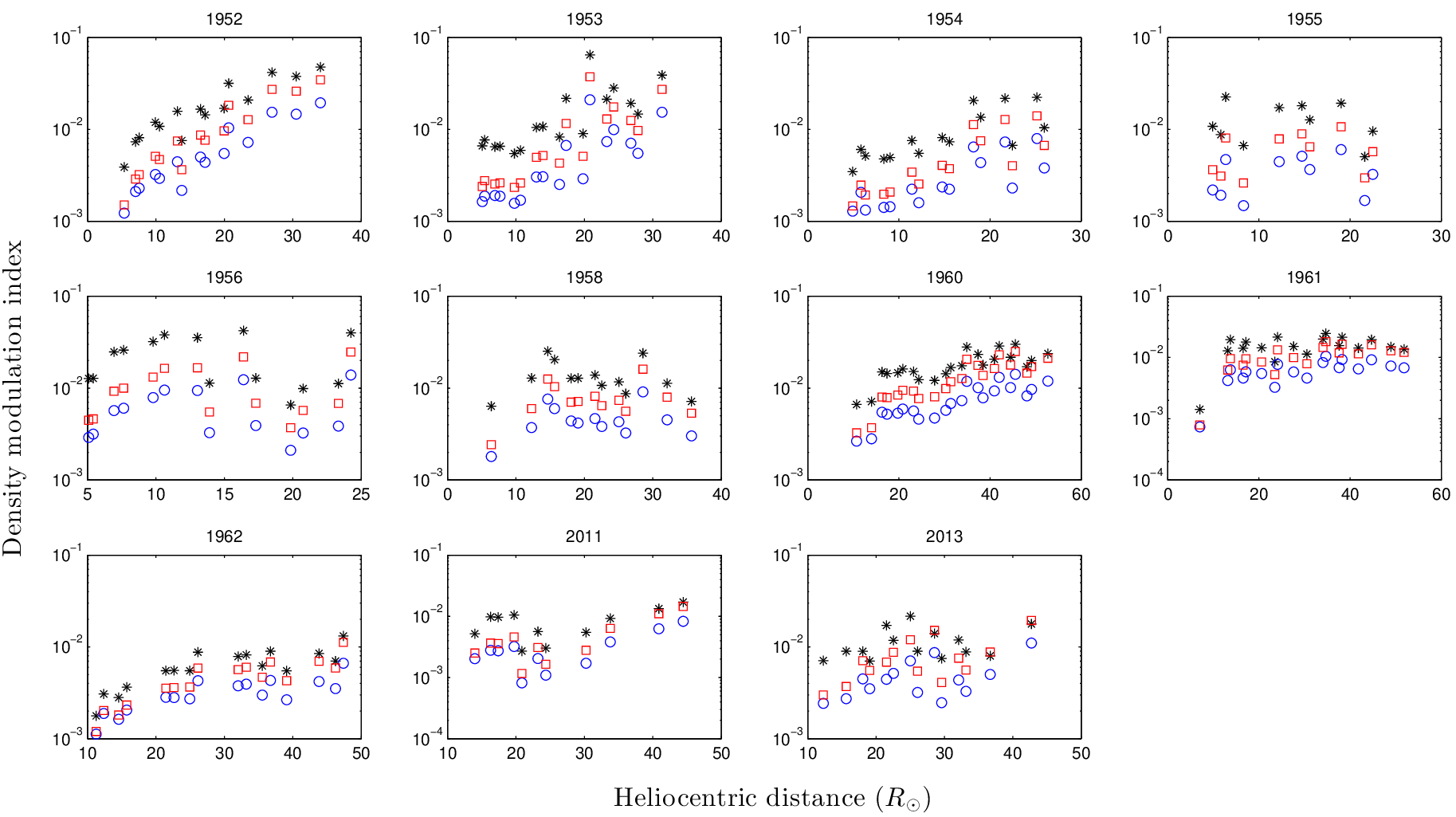}
	\caption{The Figure shows the measured density fluctuation index ($\epsilon_N$) over a heliocentric distance in different years using various inner scale models. The `*' indicates the proton inertial length model and `circles' and `squares' represent 
		the proton gyroradius model with proton temperature $10^5$ and $1.5 \times 10^6$ K respectively.}
	\label{fig:epsilon}
\end{sidewaysfigure}

Using the knowledge of $C_{N}^{2}$ the density modulation index $\epsilon_{N}$ can be estimated using Eqs~(\ref{eq:deltn}) 
and (\ref{eq:df}). \citet{Leb1998} model was used to evaluate the background solar wind density $N$.
The heliocentric distance dependence of $\epsilon_{N}$ is shown in Figure \ref{fig:epsilon} for different years. 
This quantity is computed using both the proton inertial length and proton gyroradius inner scale models. The broad conclusion 
that can be drawn is that $\epsilon_N$ ranges between 0.001 and 0.1, and its only weakly dependent on heliocentric distance.

Further, assuming the kinetic \Alfven wave dispersion equation the heating rate was derived. 

\section{Proton heating rate}

In order to estimate the heating rates in the solar wind, above mentioned density modulation indices were used. In this study, those modulation indices were used to estimate the proton heating rate ($\epsilon_{k_i}$) 
by applying the kinetic Alfvén wave dispersion relation, and compare the results with 
recent measurements obtained from angular broadening \citep{Sas2017} and interplanetary scintillation observations 
\citep{Bis14, Ing2015b}. These results were further compared with the heating rates derived from in-situ measurements 
reported by \citet{Adh2020} and \citet{Ban2020}. 

\begin{figure*}[!ht]
\centerline{\includegraphics[width=18cm]{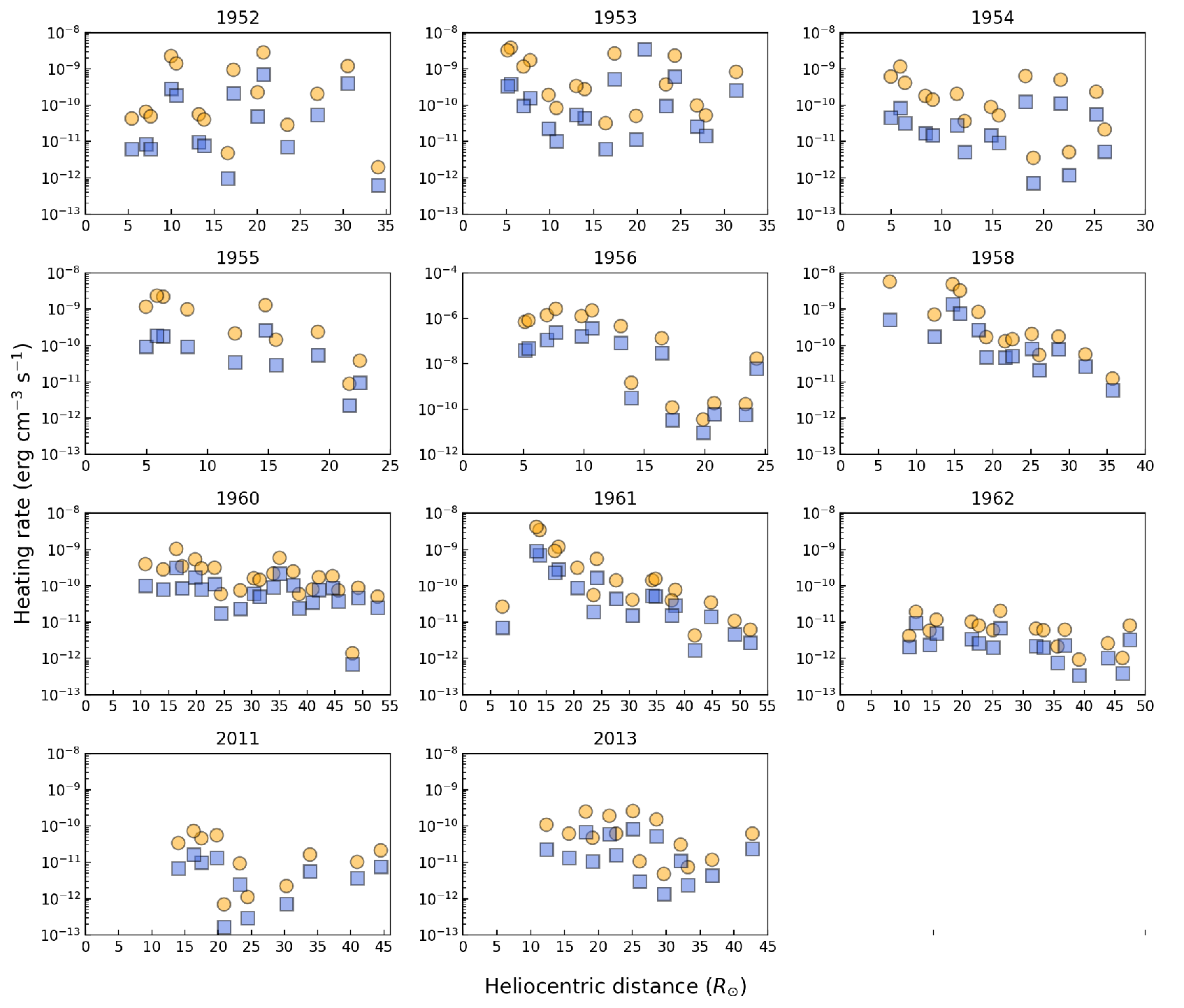}}
\caption{Heliocentric dependence of the proton heating rate in different years is shown. The markers `circle' and `square' indicate proton heating rates that are derived using the proton inertial length and proton gyroradius model, respectively. 
}
\label{fig:helio}
\end{figure*}

Density modulation indices ($\epsilon_{N_e}$) derived using the above method (see \S \ref{lab:densmod}) were used to measure the heating rates. Following \citet{Cha2009, Sas2017}, and assuming that density fluctuations at small scales are manifestations of low frequency, 
oblique ($k_{\perp} \gg k_{\parallel}$), $\rm Alfv\acute{e}n$ wave turbulence and are often referred to 
kinetic $\rm Alfv\acute{e}n$ waves. 
Here, the quantities $k_{\perp}$ 
and $k_{\parallel}$ are the components of the wave vector k
in perpendicular and parallel direction to the background large-scale magnetic field, respectively.

As previously discussed we envisage a situation where the ``balanced'' counter propagating $\rm Alfv\acute{e}n$ 
waves (i.e. with zero helicity) cascade and resonantly damps on the 
protons at the inner scale and thereby heats the solar wind. 
Because of the passive mixing of the $\rm Alfv\acute{e}n$ waves with other modes at the inner scale
our proton heating rate measurements provide an upper limit. The proton heating rate (i.e. the turbulent energy cascade rate) at inner scales is \citep{Hollweg1999, Cha2009, Ing2015b}, 

\begin{equation}\label{eq:hr}
\epsilon_{k_i}(R)=c_0 \rho_p k_i(R) \delta v_{k_i}^3(R) ~ \rm erg ~cm^{-3}~s^{-1} \, ,
\end{equation}

where, $\rho_p=m_pN_e(R)~\rm g~ cm^{-3}$ with $m_p$ is the proton mass [in grams],  $k_i=2 \pi/l_i$ and $\delta v_{k_i}$ are the wavenumber and  magnitude of turbulent velocity fluctuations at inner scales, respectively. The dimensional less quantity $c_0$ is assumed to be 0.25 \citep{How2008, Cha2009, Sas2017}. 

By knowing the $\epsilon_{N_e}$, we calculated $\delta v_{k_i}$ using the 
kinetic $\rm Alfv\acute{e}n$ wave dispersion relation \citep{How2008,Cha2009,Ing2015b, Sas2017}

\begin{eqnarray}\label{eq:rmsv}
 \delta v_{k_i}(R)=\Bigg({1+{\gamma_i k_i^2(R) \rho_i^2(R)} \over {k_i(R) l_i(R)}} \Bigg) \epsilon_{N_e} (R, k_i) v_A(R) \, .
\end{eqnarray}

where, the adiabatic index $\gamma_i$ is taken to be 1 \citep{Cha2009, Sas2017}. 

The $\rm Alfv\acute{e}n$ speed ($v_A$) in the solar wind is measured using, 

\begin{equation}\label{eq:va}
v_A(R)=2.18\times 10^{11} \mu^{-1/2} N_e^{-1/2}(R)B(R) ~\rm cm~s^{-1}, 
\end{equation}

The magnetic field strength (B) is estimated using the Parker spiral magnetic field in the ecliptic 
plane using \citep{Wil1995}, 

\begin{equation}\label{eq:parker}
B(R)= 3.4 \times 10^{-5} R^{-2} (1+R^2)^{1/2} ~ \rm Gauss, 
\end{equation}

where, `R' is the heliocentric distance in units of AU. 

\begin{figure}[!ht]
\centerline{\includegraphics[width=15cm]{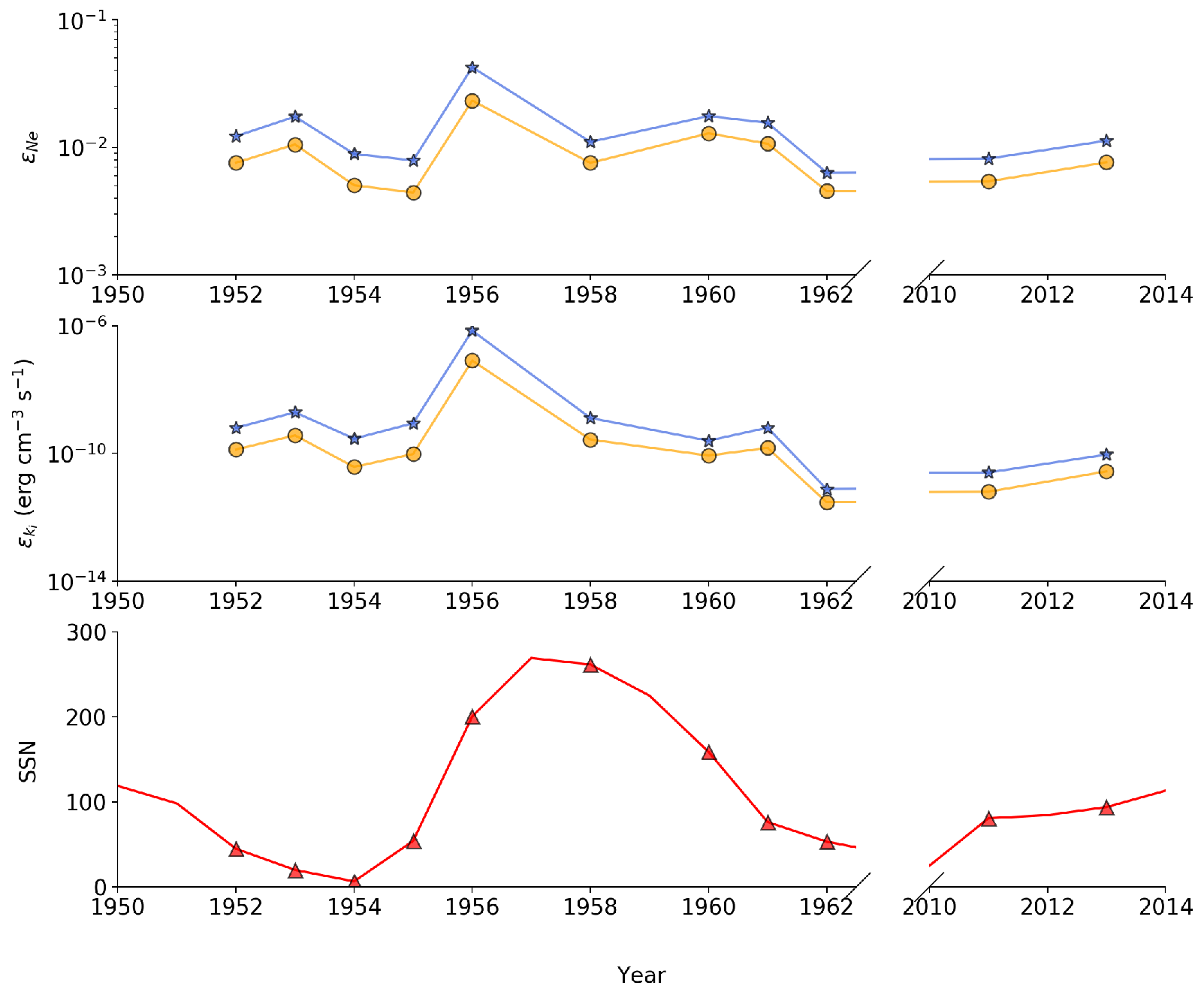}}
\caption{(top) Solar cycle dependence of the modulation index averaged over the heliocentric distance $5-45~ R_{\odot}$.
(middle) Solar cycle dependence of the proton heating rate averaged over the heliocentric distance $5-45~ R_{\odot}$. The measurements for various inner scale models, i.e., proton inertial length and proton gyroradius models are indicated with the symbols `circle' and `square', respectively. (bottom) The solid line shows the yearly averaged sunspot number \citep{Cle2016} and the symbol `triangle' indicates the sunspot number in which we have the radio observations. The Figure shows a clear solar cycle dependence of the proton heating rate.}
\label{fig:sc}
\end{figure}

The derived proton heating rates in different years are shown in Figure \ref{fig:helio} and we found that heating rates vary from $\approx 1.58 \times 10^{-14}$ to  $1.01 \times 10^{-8} ~\rm erg~ cm^{-3}~ s^{-1}$ over the heliocentric distances 5 - 45 $\rm R_{\odot}$. The markers `circle' and `square' indicate proton heating rates derived assuming different inner scale models - proton inertial length and proton gyroradius model. 

At 5 $\rm R_{\odot}$, in the coronal holes (i.e., in the fast solar wind), the estimated proton heating rates range from $2 \times 10^{-10}$ and $1.4 \times 10^{-8} ~\rm~ erg~ cm^{-3} ~s^{-1}$ \citep{Cha2009}. Similarly, at 1 AU the estimated heating rate is $5 \times 10^{-16} \rm~ erg ~cm^{-3}~ s^{-1}$ \citep{Cha2009}. The heating rates derived assuming density fluctuations are due to the kinetic \Alfven waves in the heliocentric distance range 2-174 $\rm R_{\odot}$ using interplanetary scintillation (IPS) observations \citep{Hew1963,Man2000,Jan11,Sas2019b} are $3 \times 10^{-8} ~\rm erg~ cm^{-3}~ s^{-1}$ (during solar maximum) and $\approx 10^{-15} ~\rm erg~ cm^{-3}~ s^{-1}$ (during solar minimum) consistent with our estimates \citep{Ing2015b}. Using two-dimensional imaging angular broadening observations of the Crab Nebula, the measured heating rates are varied from $2.2 \times 10^{-13}$ to $ 1.0 \times 10^{-11} ~\rm erg~ cm^{-3}~ s^{-1}$ in the heliocentric distance range $9 - 20~ R_{\odot}$ \citep{Sas2017}. The recently reported heating rates in the heliocentric distance range 1.5 - 4.0 $R_{\odot}$ varied from $\approx 3.31 \times 10^{-10}$ to $4.5 \times 10^{-7}\rm~ erg~ cm^{-3}~ s^{-1}$ \citep{Ste2020}. Further, the author extrapolated these heating rates to the distances 0.3-0.6 AU which range from $\approx 10^{-15}$ to $10^{-14}\rm~ erg~ cm^{-3}~ s^{-1}$ and at 1 AU, the extrapolated heating rates are few times of $10^{-16}\rm~ erg~ cm^{-3}~ s^{-1}$.

Using in-situ measurements by Parker Solar Probe, \citet{Ban2020} estimated energy transfer rates of $8.7 \pm 0.3 \times 10^{-13}~\rm erg~ cm^{-3}~ s^{-1}$ at 36 $R_{\odot}$ and $5.8 \pm 1.3 \times 10^{-14} ~\rm erg~ cm^{-3}~ s^{-1}$ at 54 $R_{\odot}$. They originally have quoted numbers in units of ${\rm J\,kg^{-1}\,s^{-1}}$. We have multiplied their numbers by $ N_{p}\, m_{p}$ (where $N_{p}$ is the solar wind density derived using Leblanc model \citep{Leb1998} and $m_{p}$ is the proton mass) to arrive at heating rates in units of ${\rm erg\,cm^{-3}\,s^{-1}}$. By comparison, the proton heating rate at 36 $R_{\odot}$ from our results (see Figure \ref{fig:helio}) range from $\approx 2.8 \times 10^{-10}$ to $\approx 7.4 \times 10^{-13}$ ${\rm erg\,cm^{-3}\,s^{-1}}$. 
Similarly, \citet{Adh2020} reported that heating rates in the heliocentric distance $\approx 1.6 - 100 ~R_{\odot}$ range from $1.06 \times 10^{-4}~\rm~to~ 1.73 \times 10^{-14}~ erg~ cm^{-3}~ s^{-1}$. Authors also reported that the heating rate due to the nearly in-compressible/slab turbulence in the heliocentric distance $\approx 1.3 - 100 ~R_{\odot}$ range from $4.24 \times 10^{-7}~\rm~to~ 1.11 \times 10^{-14}~ erg~ cm^{-3}~ s^{-1}$. A summary of these proton heating rates is given in Table \ref{tab:hr123}.

\begin{table}[h!]
\centering
 \begin{tabular}{c c c c}
 	\hline
 	S.No & R             & Proton heating rate                             & References       \\
 	     & ($R_{\odot}$) & ($\rm erg ~cm^{-3}~ s^{-1}$)                    &                  \\ \hline
 	                           \multicolumn{4}{c}{Remote sensing}                             \\ \hline
 	 1   & 5 - 45        & $1.58 \times 10^{-14}$ -  $1.01 \times 10^{-8}$ & Present work     \\
 	 2   & 5             & $2 \times 10^{-10}$ - $1.4 \times 10^{-8}$      & \citet{Cha2009}  \\
 	 3   & 215           & $5 \times 10^{-16}$                             & \citet{Cha2009}  \\
 	 4   & 2  -174       & $3 \times 10^{-8}$ - $10^{-15}$                 & \citet{Ing2015b} \\
 	 5   & 9 - 20        & $2.2 \times 10^{-13}$ - $ 1.0 \times 10^{-11}$  & \citet{Sas2017}  \\
 	 6   & 1.5 - 4.0     & $3.31 \times 10^{-10}$ - $4.5 \times 10^{-7}$   & \citet{Ste2020}  \\
 	 7   & 64.5 - 129    & $10^{-15}$ - $10^{-14}$                         & \citet{Ste2020}  \\
 	 8   & 215           & $10^{-16}$                                      & \citet{Ste2020}  \\ \hline
 	                               \multicolumn{4}{c}{In-situ}                                \\ \hline
 	 9   & 36            & $8.7 \pm 0.3 \times 10^{-13}$                   & \citet{Ban2020}  \\
 	 10  & 54            & $5.8 \pm 1.3 \times 10^{-14} $                  & \citet{Ban2020}  \\
 	 11  & 1.6 - 100     & $1.06 \times 10^{-4}$ - $1.73 \times 10^{-14}$  & \citet{Adh2020}  \\
 	 12  & 1.3 - 100     & $4.24 \times 10^{-7}$ - $1.11 \times 10^{-14}$  & \citet{Adh2020}  \\ \hline
 \end{tabular}
\caption{Summary of proton heating rates in the solar wind}
\label{tab:hr123}
\end{table}

As the density modulation indices (see Figure \ref{fig:epsilon}) and heating rates (see Figure \ref{fig:helio}) are weakly dependent with heliocentric distance, we averaged the observations that are carried out in different years and plotted in Figure \ref{fig:sc}. The upper and middle panels of Figure \ref{fig:sc} are the averaged density modulation indices and proton heating rates for different inner scale models, respectively. The lower panel shows the yearly averaged sunspot number. Figure \ref{fig:sc} shows that the derived density modulation indices and heating rates closely follow the solar cycle. During solar maximum, the slow solar wind drives in all the directions and hence \citet{Sas2016} had justified the lower modulation index in 1958 (also refer to upper panel of Figure \ref{fig:sc}). Following the lower density modulation indices, heating rates are lower during the solar maximum.

\section{Solar cycle dependence of modulation index and proton heating rate}

Since the heliocentric distance dependence of $\epsilon_N$ is rather weak, it is meaningful to compute an average for this quantity for each year.
The average of $\epsilon_N$ between 10 and 45 $R_{\odot}$ is plotted as a function of time in the upper panel of Figure \ref{fig:epsilon}. 
Comparison with the lower panel, which shows the sunspot numbers, shows that $\epsilon_N$ broadly follows the solar cycle. 

However, we note that $\epsilon_N$ shows a prominent dip around 1958, which happens to be the year with the highest sunspot number of 
the data we have examined. Although the dip comprises only one data point, the following could be a tentative explanation for it:
\citet{Cel87} notes that the modulation index ($\epsilon_N$) is positively correlated with the temperature of solar wind protons. At 1 AU, 
it has been observed that the proton temperature is positively correlated with solar wind speed \citep{Lop86}.
Taken together, this implies that $\epsilon_N$ should be larger in the fast solar wind than in the slow solar wind. 
During the solar minimum, the Sun's large-scale magnetic field is predominantly dipolar. Consequently, higher latitudes 
are dominated by fast ($\approx 700 ~km/s$) solar wind emanating from coronal holes. Lower latitudes, on the other hand, 
are dominated by the slow solar wind ($\approx 400 ~km/s$) emanating from near the streamer belt. During solar maximum, however,
the large-scale solar magnetic fields is multi polar. Coronal holes are not as prevalent and slow solar wind is observed over all 
heliolatitudes \citep{Mcc00, Asa98}. Since 1958 was associated with a high sunspot number (the highest of the years we have 
considered), we expect slow solar wind (and low proton temperatures) at all heliolatitudes because the magnetic field is multipolar. 

Furthermore, \citet{Asa98} suggest that the modulation index of the high speed solar wind (which is usually observed near solar minimum) 
shows significant evolution with heliocentric distance. Our results show that the modulation index does not vary 
appreciably with heliocentric distance during the solar maximum years of 1956, 1958, 1960, 1961 and 2013, when the slow solar wind is expected 
to dominate. Our results are thus consistent with the converse of the conclusions reached by \citet{Asa98}.

\section{Dissipation scales}

\subsection{Estimating the dissipation scale}\label{sec:sas1}

The Crab nebula images can be constructed by suitably inverse Fourier transforming a set of calibrated visibilities using the Astronomical Image Processing System (AIPS)\footnote{http://www.aips.nrao.edu/index.shtml}. On the other hand, the theoretical view of the structure function relates it to the spectrum of turbulent fluctuations that give rise to the observed scatter broadening. The general structure function (\citep{Ing2015}) is used to find the dissipation scales. The GSF is, 

\begin{equation}
\begin{split}
D_{\rm th}(s) & = C \times l_i(R)^{\alpha-2} \\
	   & {\times \bigg\{ { _1F_1} {\bigg[ - {{\alpha-2} \over 2},~1,~ - \bigg( {s \over l_i(R)} \bigg)^2 \bigg]} -1 \bigg\}} \, ,
\end{split}	    
\label{the_structfun}
\end{equation}

where
\begin{equation}\label{eq:gsf1}
C = {{8 \pi^2 r_e^2 \lambda^2 \Delta L} \over {\rho~2^{\alpha-2}(\alpha-2)}} {\Gamma \big( 1 - {{\alpha-2} \over 2} \big)} 
	    {{C_N^2 (R)} \over {(1 - f_p^2 (R) / f^2)}} \, ,
\end{equation}

\begin{figure}[!ht]
\centerline{\includegraphics[width=9.5cm]{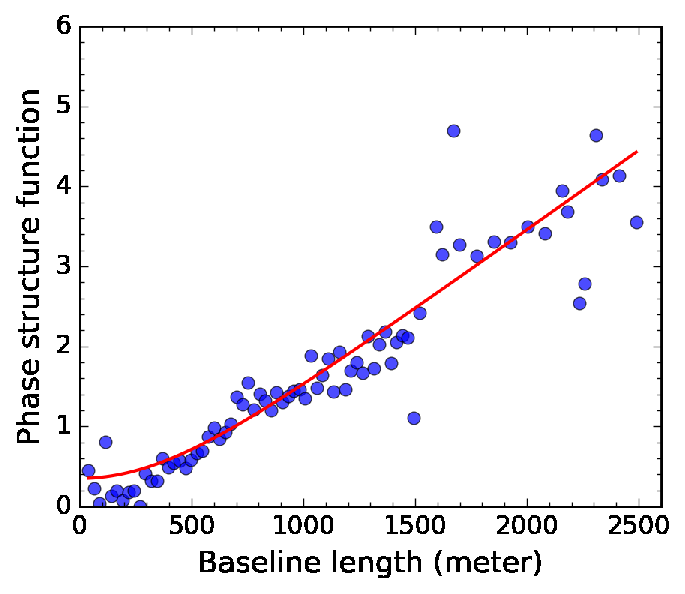}}
\caption{The circles indicate $D_{\rm obs}(s)/D_{\rm obs}(s_r)$ for the Crab Nebula observed on 18 June 2016, the radio image. The solid line represents $D_{\rm th}(s)/D_{\rm th}(s_r)$ (Eq~ \ref{the_structfun}) for $\alpha=3$ and the dissipation scale $l_i= 560$ m. The average rms error between the model fit (solid line) and the visibility data (circles) is 3.67\%}
\label{fig:fit}
\end{figure}

${ _1F_1}$ is the confluent hypergeometric function, $r_e$ is the classical electron radius, 
$\lambda$ is the observing wavelength, $R$ is the heliocentric distance, $\Delta L$ is the thickness of the scattering medium, $\rho$ is the axial ratio of the observed image, 
$f_p$ and f are the plasma and observing frequencies respectively. A suitable reference baseline $s_r$ to define a function $D_{\rm th}(s)/D_{\rm th}(s_r)$ is selected. For a given value of the inner scale $l_i$, this is a function only of the baseline $s$.
The value of $l_{i}$ that minimizes the least squared difference between $D_{\rm th}(s)/D_{\rm th}(s_r)$ (Eq~\ref{the_structfun}) and $D_{\rm obs}(s)/D_{\rm obs}(s_r)$ is regarded as our estimate for the dissipation scale. A successful fit for the observation is shown in Figure \ref{fig:fit}. Results for all our observations are listed in Table \ref{tab:one}. The root mean square percentage difference between $D_{\rm th}(s)/D_{\rm th}(s_r)$ and $D_{\rm obs}(s)/D_{\rm obs}(s_r)$ for the estimated $l_{i}$ is listed. The red circles in figure~\ref{fig:is} depict the estimated dissipation scale as a function of heliocentric distance and the error bars on the red circles correspond to the fitting errors listed in a Table~\ref{tab:one}. 

Similar attempts have been made in earlier studies. For instance, \citet{Col1989} derived structure functions from radar observations, while \citet{Ana1994} used those obtained from imaging observations. However, both studies estimated the dissipation scale ($l_i$) by identifying a break between the asymptotic predictions for the $s \ll l_{i}$ and $s \gg l_{i}$ regimes of the structure function. In contrast, our approach involves fitting the generalized structure function, which incorporates both the asymptotic regimes ($s \ll l_{i}$ and $s \gg l_{i}$) as well as the intermediate transition region ($s \approx l_{i}$), directly to the observed visibility data with $l_{i}$ treated as a fitting parameter. This approach yields a significantly more accurate estimate of the dissipation scale. 

\subsection{Comparing with dissipation scale models}\label{sec:sas2}

Most theoretical studies of solar wind turbulence particularly those that attempt to describe the kinetic regime suggest that the turbulent spectrum steepens at what is commonly referred to as the “proton scale.” This scale is interpreted differently in the literature: some associate it with the proton inertial length, others with the proton gyroradius (e.g., \citet{Bol2015}), and still others propose that dissipation occurs over a range of scales \citep{Tol2015}. The proton inertial length model \citep[e.g.,][]{Col1989, Yam1998, Bru2014} envisions resonant damping of protons on Alfvén waves, leading to the following expression for the dissipation scale:

\begin{equation}
 l_{i}(R) = d_i \equiv V_{A}/\Omega_{p} = 228~N_e(R)^{-1/2} \,\, {\rm km}\, \, ,
 \label{eq:inertiallength} 
\end{equation}
where $V_{A}$ is the $\rm Alf\acute{v}en$ speed, $\Omega_{p}$ is the proton cyclotron frequency and $N_{e}(R)$ is the electron number density in ${\rm cm}^{-3}$ at a heliocentric distance $R$. The solid line in Figure~\ref{fig:is} depicts the prediction of the proton inertial length model with the electron density given by the model of Leblanc \citep{Leb1998}. The proton gyroradius is given by 
\begin{equation}
l_i (R) = 102 \mu^{1/2} {T_{i}(R)}^{1/2} {B(R)}^{-1} ~\rm cm,
\label{eq:protongyroradius}
\end{equation}
where $\mu$ is the ratio of mass of ion to mass of proton (taken to be 1 for our purposes), $T_i$ is the proton temperature in eV and B is the magnetic field in G. Note that two models for the heliospheric magnetic fields were used: one for a nominal Parker spiral magnetic field in the ecliptic e.g., \citep{Wil95},

\begin{equation}\label{eq:parker}
B(R)= 3.4 \times 10^{-5} R^{-2} (1+R^2)^{1/2} ~ \rm Gauss, \, ,
\end{equation}
where $R$ is the heliocentric distance in units of AU (1 AU = $215 R_{\odot}$),
and one using extrapolations using Helios observations \citep{Ven2018}, 

\begin{equation}
B (R) = 1.089 (0.0131\times SSN + 4.29)\times R^{-1.66} ~\rm nT \, ,
\label{eq:Bothmer}
\end{equation}

where $SSN$ refers to the sunspot number on the day of observation. The dissipation scale as a function of heliocentric distance, using the proton gyroradius prescription of Eqs.~(\ref{eq:protongyroradius}) and (\ref{eq:parker}) with a proton temperature of $10^{5}$~K, is depicted by the red dashed line in Figure~\ref{fig:is}, while that using the prescription of Eqs.~(\ref{eq:protongyroradius}) and (\ref{eq:Bothmer}) with a proton temperature of $10^{6}$~K is depicted by the blue dash-dotted line in Figure~\ref{fig:is}. Except for the first point (at 2.2~$R_{\odot}$), our estimates of the dissipation scale are more consistent with the proton gyroradius interpretation than with that invoking the proton inertial length. The estimated value of the dissipation scale for the VLA observation of 1430–155 at 2.2~$R_{\odot}$ is substantially larger than those for the GRAPH observations, which were taken at larger heliocentric distances (Table~\ref{tab:one}). There are three points to note in this connection. First, the sunspot number corresponding to the day of the VLA observation was larger than those for the GRAPH observations (Table~\ref{tab:one}). Second, the source 1430–155 was located much closer to the solar south pole than the Crab Nebula, suggesting that radiation from 1430–155 predominantly sampled turbulence in the fast solar wind, while the Crab Nebula observations sampled mostly slow solar wind turbulence. Finally, the short-baseline coverage (ranging from a few hundred meters to a few kilometers) for the VLA A-array, with which the observations of 1430–155 were made, is much sparser than that for the GRAPH. The relative lack of data points at short spacing tends to bias the model-fitting procedure toward higher values of $l_i$. Therefore, SKA-low and SKA-mid play a crucial role in full fill the requirement of the short baselines so that we can precisely find the inner scale lengths. 

\begin{table}
\centering
\vspace*{5px}
\begin{tabular}{|c|c|c|c|c|c|c|c|c| }
        \hline \hline

 S.&   Date & Source & R &  & Phase of the & $l_i$ & RMSE \\
 No. &   & Name   & ($R_{\odot}$) & SSN  & solar cycle&  (meter) & (\%)  \\
 (1) & (2) & (3) & (4) & (5) & (6) & (7) & (8) \\
    \hline
   1 &  1988 Nov 02 & 1430-155  & 02.5    & 143 & Ascending      & 13500   & 2.7  \\
   2 &  2016 Jun 18 & Crab      & 13.46   & 47  & Descending     & 560 & 3.67 \\
   3 &  2016 Jun 19 & Crab      & 16.83   & 50  & Descending     & 600 & 2.34 \\\
   4 &  2016 Jun 20 & Crab      & 20.27   & 36  & Descending     & 770 & 3.53 \\
   5 &  2017 Jun 10 & Crab      & 17.68   & 0   & Descending     & 590 & 2.0  \\
   6 &  2017 Jun 12 & Crab      & 10.97   & 0   & Descending     & 520 & 5.1 \\
   
       \hline \hline

\end{tabular}
\caption{Column 2 lists the date on which the scatter-broadened source was observed. Column 3 lists the source name and column 4 shows the projected heliocentric distance in units of the solar radius. Column 5 and 6 list the total sunspot number and phase of the solar cycle on that day. Column 7 lists the estimated dissipation scale length in meters while column 8 lists the percentage root mean square error (RMSE) associated with this estimate. 2017 Jun 10 and 12 were sun-spotless days, hence the zeroes corresponding to those days in column 5.}
\label{tab:one}
\end{table}

\begin{figure}[!ht]
\centerline{\includegraphics[width=9.5cm]{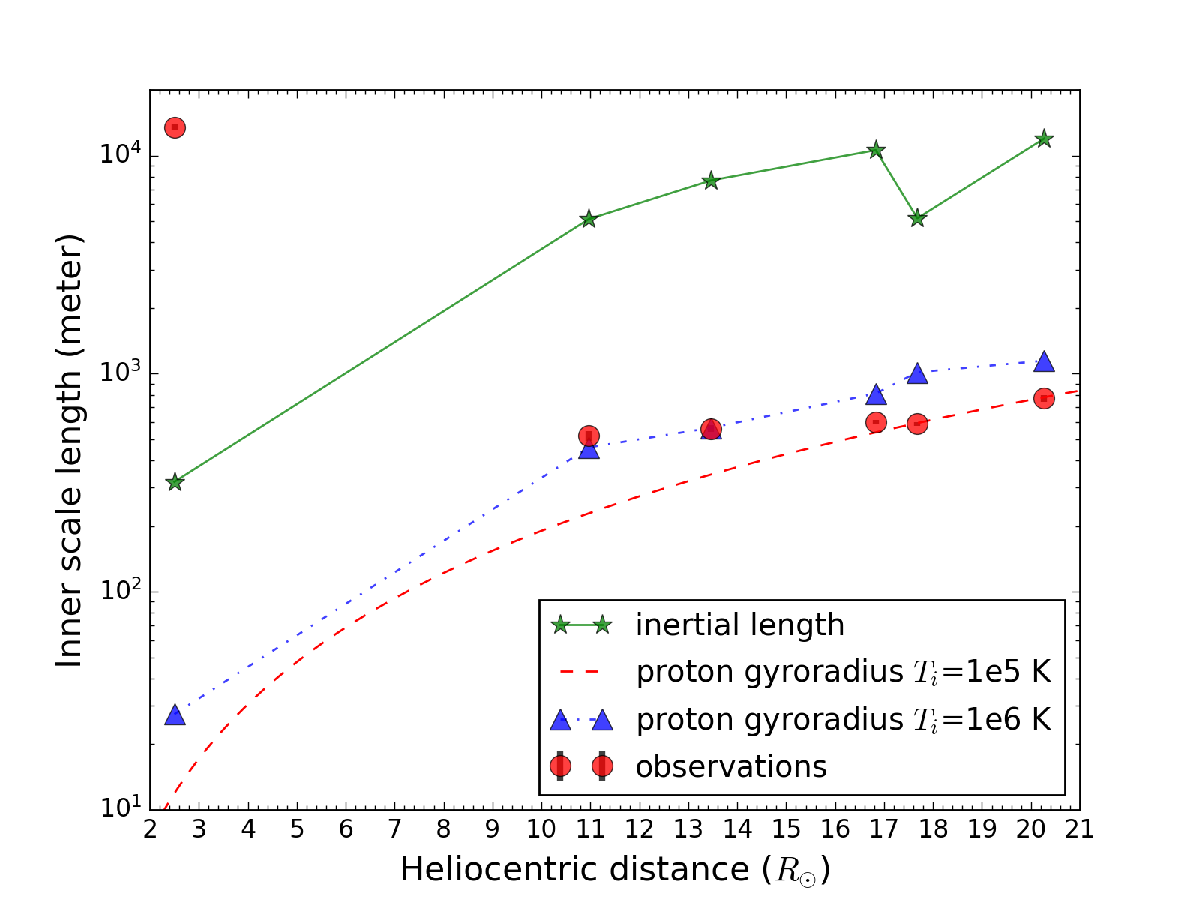}}
\caption{The dissipation length ($l_i$) of density turbulence 
in the solar wind as a function of projected heliocentric distance in units of $R_{\odot}$. The red circles indicates the dissipation scale lengths derived from the observations. The solid line indicates the prediction of proton inertial length model. The dashed line indicates the proton gyroradius computed using a temperature of $10^{5}$ K and the Parker spiral magnetic field model \citep{Wil95}, while the dash-dotted line indicates the proton gyroradius using a temperature of $10^{6}$ K and the magnetic field model of Venzmer \& Bothmer \citep{Ven2018}.
}
\label{fig:is}
\end{figure}

\section{Observations through streamers}\label{observations}

The radio observations were carried out with the Gauribidanur Radioheliograph (GRAPH; \citet{Ram98, Ramesh2011}) at 80~MHz during the local meridian transit of the Crab Nebula. The GRAPH is a T-shaped interferometric array with baselines ranging from approximately 80 to 2600~m. It provides an angular resolution of about 5~arcmin at 80~MHz, and a minimum detectable flux density of roughly 50~Jy (at the 5$\sigma$ level) for an integration time of 1~s and a bandwidth of 1~MHz. Cygnus~A, with a flux density of approximately 16,296~Jy at 80~MHz, was used for calibration. The flux density of the Crab Nebula when observed far from the Sun and thus unaffected by coronal scattering is about 2015~Jy at 80~MHz. The Crab Nebula was imaged at different projected heliocentric distances, listed in column~(3) of Table~\ref{tab:table-1}, during the years 2016 and 2017.

\begin{figure*}[!ht]
\begin{minipage}{0.4\textwidth}
\centerline{\includegraphics[width=1.4\textwidth]{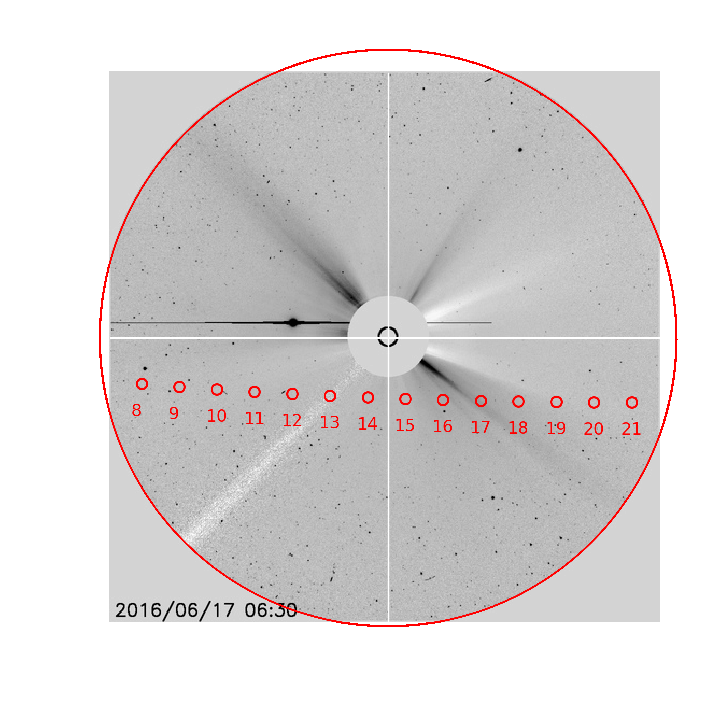}}
\end{minipage}
\begin{minipage}{0.6\textwidth}
\centering
\includegraphics[width=0.75\textwidth]{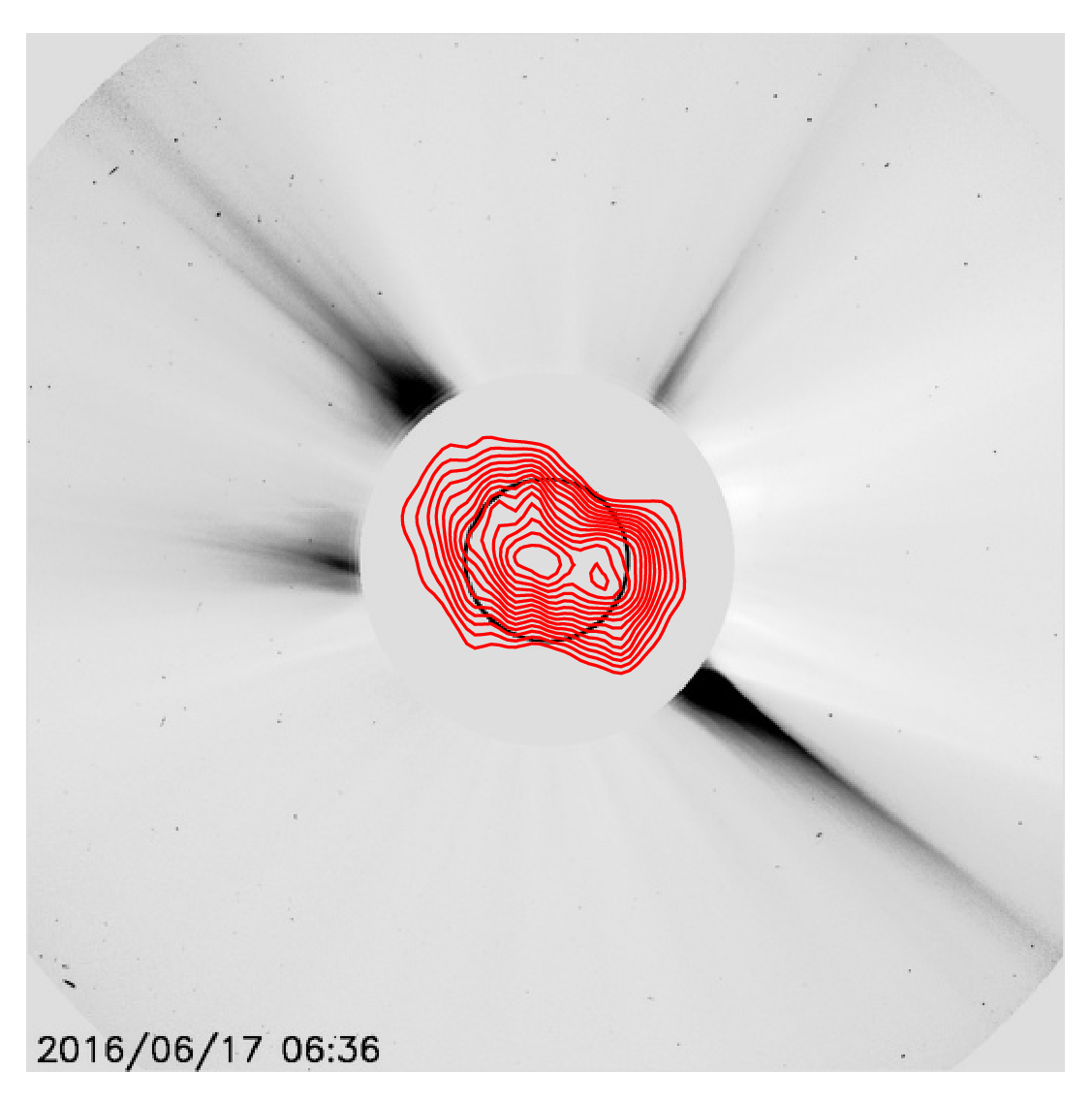}
\end{minipage}
\caption{The left panel shows the SOHO/LASCO C3 image of the solar corona (inverted grey scale image) observed on 17 June 2016 at 06:30 UT. 
The innermost black circle depicts the solar disk (radius $=1~R_{\odot}$). 
The next concentric circle is the occulting disk of the coronagraph and its radius is $3.5~R_{\odot}$. 
The outermost circle marks a heliocentric distance of $30~R_{\odot}$. In both the images, the black features are coronal streamers. Solar north is up 
and east is to the left.
The small circles superposed on the image represent the location of the Crab nebula on 
different days during the period 8 June 2016 to 21 June 2016. 
Its closest approach to the Sun is on 14 June 2016 at a heliocentric distance of $\approx 5~R_{\odot}$.
The coronal streamer in the south-west quadrant occults the Crab nebula on 17 June 2016 
at a projected heliocentric distance $\approx 10.2~R_{\odot}$. 
The position angle (PA, measured counterclockwise from north) of the streamer is $\approx 235^{\circ}$.
Right panel shows the SOHO/LASCO C2 image of the solar corona (inverted grey scale) on 17 June 2016 at 06:36 UT.
The red contours represent observations of the solar corona using the GRAPH at 80 MHz. The elongated radio contours
correspond to emission from the streamers.}
\label{fig:lasco}
\end{figure*}

White-light images of the solar corona obtained with the Large Angle and Spectrometric Coronagraph (LASCO) onboard the Solar and Heliospheric Observatory (SOHO; \citealt{Bru95}) were used to provide general context and to identify large-scale coronal features such as streamers. Figure \ref{fig:lasco} displays LASCO C3 (left) and C2 (right) images of the solar corona observed on 17 June 2016. The dark structures in both inverted grayscale images correspond to coronal streamers. The apparent positions of the Crab Nebula between 8 and 21 June 2016 are indicated by red circles on the LASCO C3 images. On 17 June 2016, the Crab Nebula was seen through a streamer located in the south-west quadrant. This streamer was associated with the active region NOAA 12555, centered at heliographic coordinates S09W71. The contours overlaid on the LASCO C2 image represent 80 MHz radio emission from the Gauribidanur Radioheliograph, showing enhanced emission from streamers in both the north-east and south-west quadrants.

\begin{figure*}[!ht]
\centering
\begin{tabular}{cccc}
\includegraphics[width=.45\textwidth]{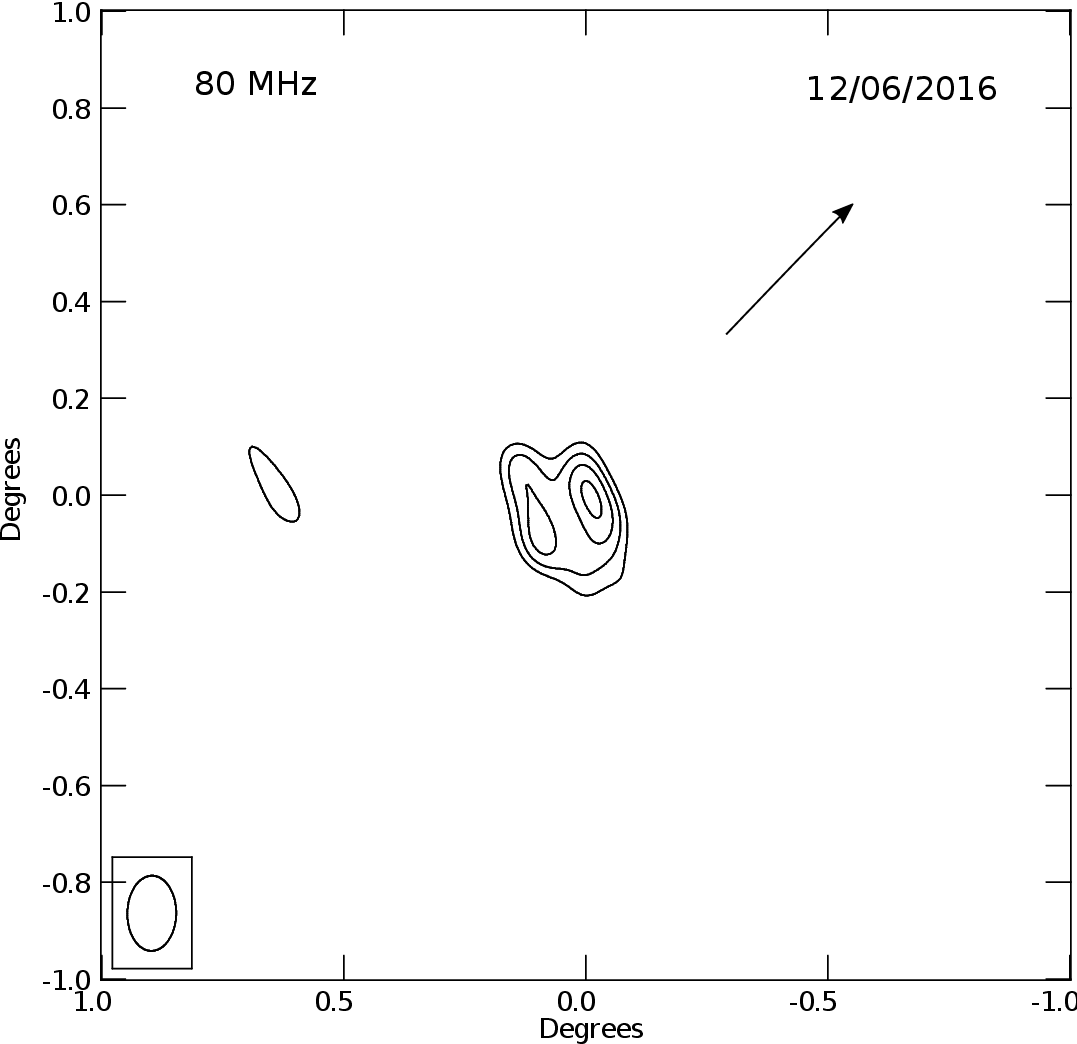} &
\includegraphics[width=.45\textwidth]{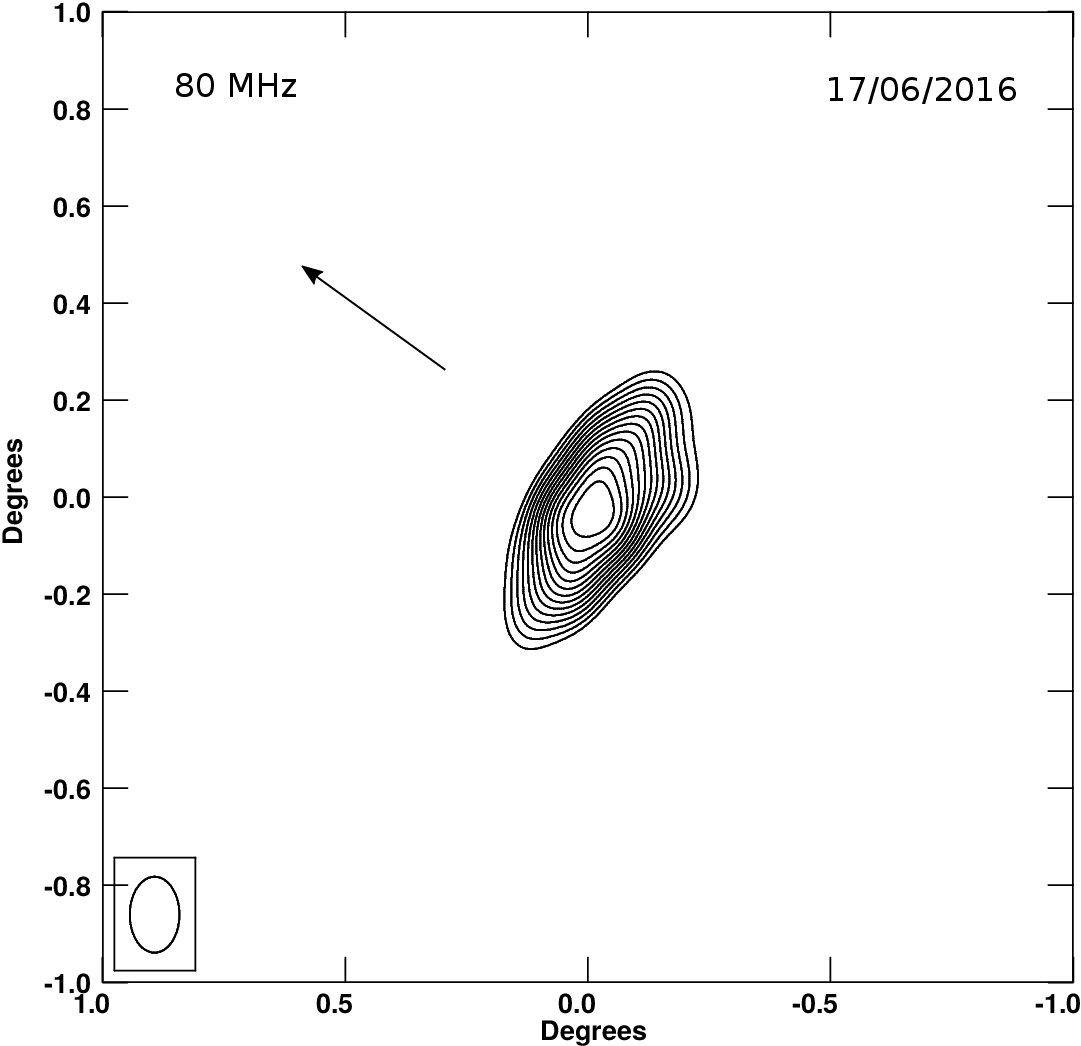} \\
\includegraphics[width=.45\textwidth]{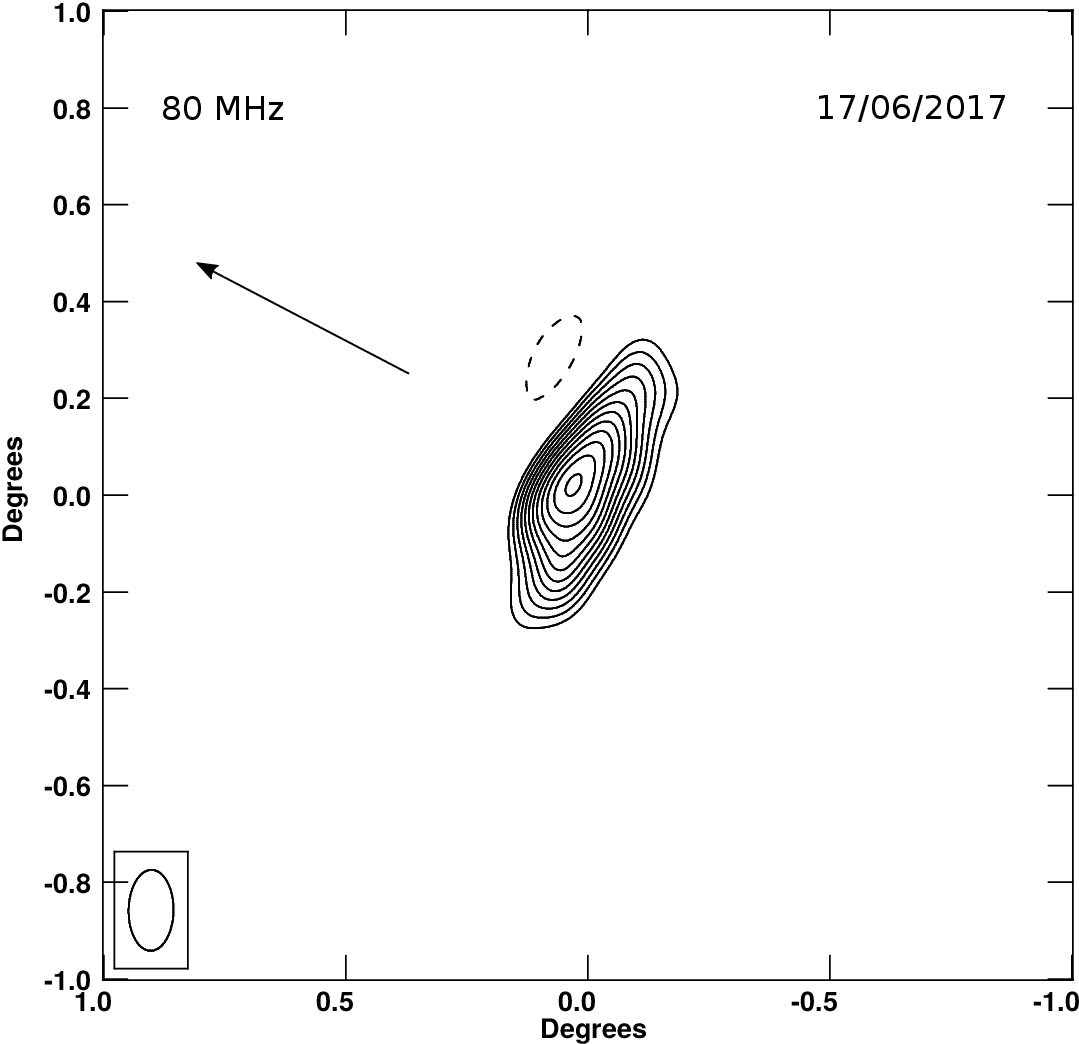} &
\includegraphics[width=.45\textwidth]{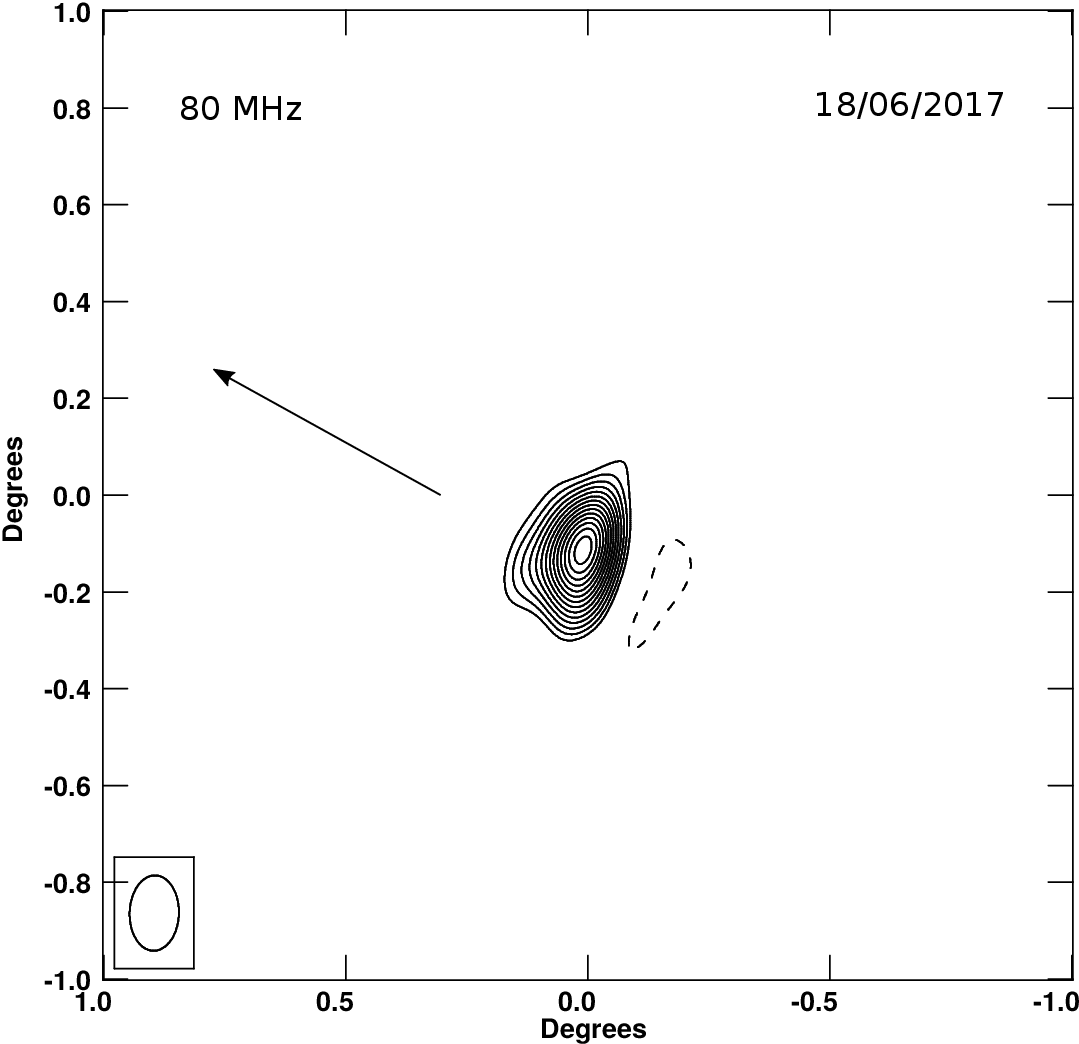}  \\ 
\end{tabular}
\caption{The image on 12 June 2016 shows the angular broadened Crab nebula at a heliocentric distance of $10.18~ R_{\odot}$ during its ingress into the inner solar wind.
The images on 17 June 2016 (at $10.2~ R_{\odot}$), 17 ($9.41~ R_{\odot}$) and 18 June 2017 ($12.61~ R_{\odot}$) depict the angular broadened Crab nebula 
observed through coronal streamers during its egress from the solar wind. The arrows depict the sunward direction on each day. The major axis of each 
image is perpendicular to the magnetic field lines, which are directed radially outward from the Sun.}
\label{fig:graph_images}
\end{figure*}

Representative 80 MHz GRAPH images of the Crab Nebula are shown in Figure \ref{fig:graph_images}. The image from 12 June 2016 corresponds to an ingress observation through the solar wind at a heliocentric distance of $10.18~R_{\odot}$. The image on 17 June 2016 was obtained at $10.20~R_{\odot}$, while those on 17 and 18 June 2017 correspond to $9.41~R_{\odot}$ and $12.61~R_{\odot}$ during egress, respectively. On 17 June 2016 and again on 17–18 June 2017, the Crab Nebula was occulted by coronal streamers. These scatter-broadened images exhibit pronounced anisotropy—a feature previously reported for the Crab Nebula \citep{Ble1972, Den1972} and other compact radio sources \citep{Arm90, Ana1994}. The major axis of the broadened images is consistently oriented perpendicular to the heliocentric radial direction, which is generally assumed to trace the magnetic field direction at these heliocentric distances. The parameters corresponding to all Crab Nebula observations in 2016 and 2017 are summarized in Table \ref{tab:table-1}.

\begin{figure}[!ht]
\centerline{\includegraphics[width=12cm]{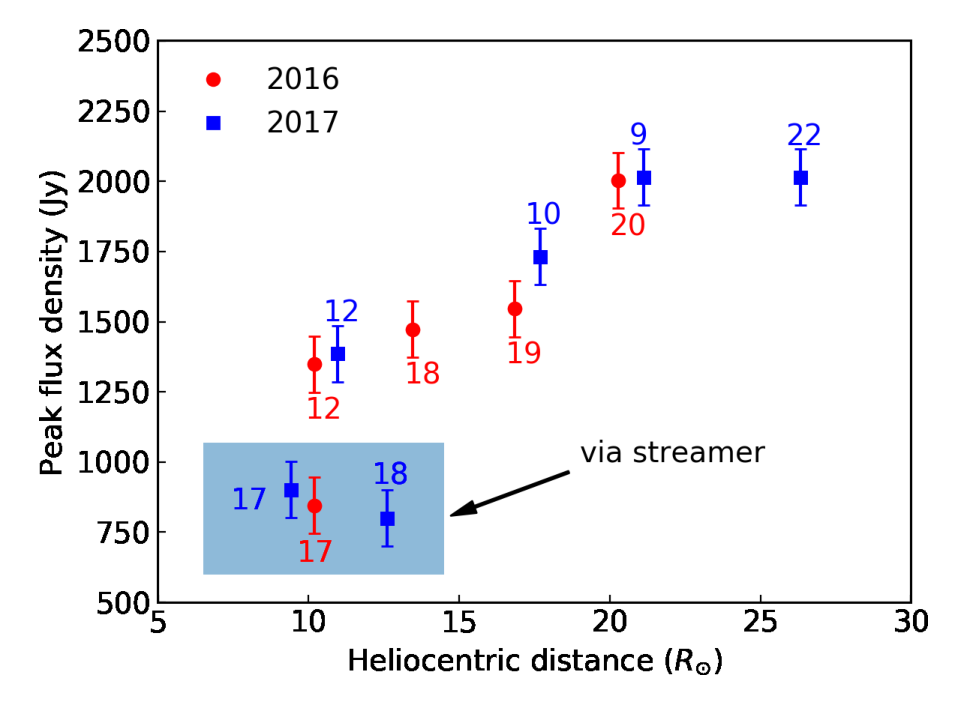}}
\caption{Peak flux density of the Crab nebula on different days of June 2016 is shown in red circles and 2017 in blue squares. 
The red and blue data points shown 
in the shaded area show instances when the Crab nebula was observed through a streamer in 
2016 and 2017 respectively. 
} 
\label{fig:peak}
\end{figure}

Figure \ref{fig:peak} indicate the observed peak flux density of the Crab nebula with respect to its projected heliocentric distance. 
The red circles and blue squares are for the 2016 and 2017 observations respectively.
Note that, in a given year the data points obtained during ingress and egress were plotted together with the (projected) heliocentric distance.

The observations within the shaded region in Figure \ref{fig:peak} correspond to periods when the Crab nebula was occulted by a coronal streamer. The peak flux density during these instances is noticeably lower than that observed at similar heliocentric distances when the Crab nebula was not occulted by a streamer. This reduction can be attributed to the line of sight intersecting a greater amount of coronal plasma during streamer occultation, resulting in enhanced scatter broadening. Consequently, the source appears more extended, leading to a decrease in the observed peak flux density.

\begin{figure}[!ht]
\centerline{\includegraphics[width=16cm]{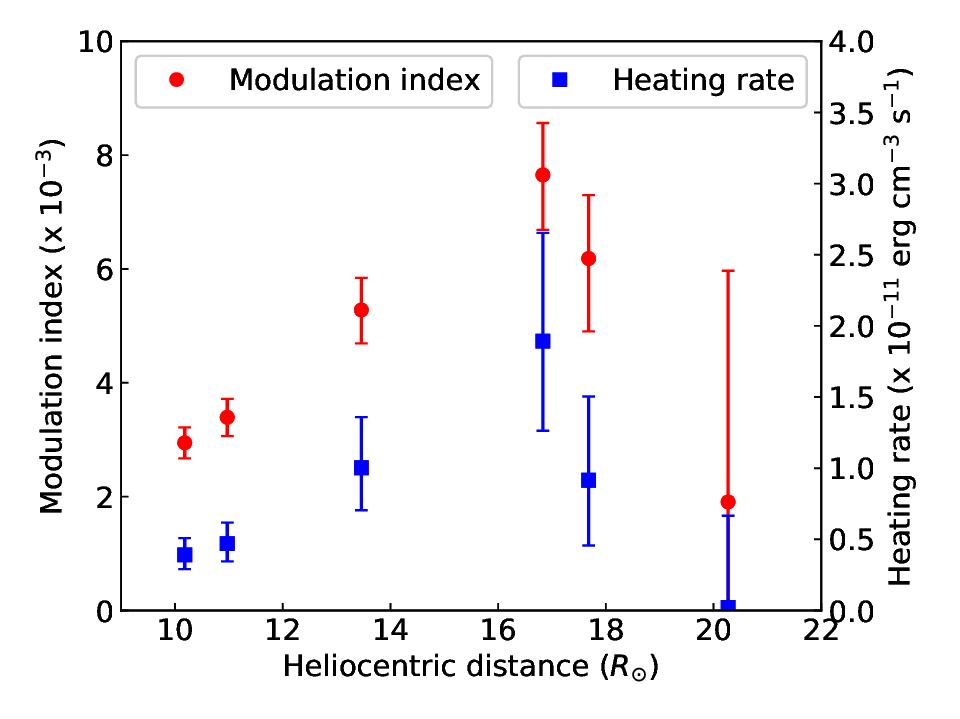}}
\caption{The variation of the density modulation index (red circles) and the solar wind 
proton heating rate (blue squares) with projected heliocentric distance. We note that the proton heating rate is 
correlated with the density modulation index.} 
\label{fig:hr}
\end{figure}

\begin{table}
\centering
\vspace*{5px}

 \begin{tabular}{|c|c |c| c| c| c| c| c| c| c| c| c| c| c| c| c| c|} 
 \hline
 \hline
 S.No & Date & R & Peak flux density& $\rm \rho $ & $\epsilon_{N_e}$ & Heating rate \\
   &  & $\rm (R_{\odot})$ & (Jy) & & & ($\rm erg~ cm^{-3}~ s^{-1}$) \\
 (1)  & (2) & (3) & (4) & (5) & (6) & (7) \\
  
   \cline{1-7}
    \multicolumn{7}{|c|}{Line of sight to the Crab does not include a streamer} \\
        \cline{1-7}

1 &	12 June 2016 &	10.18 &	1349 &	1.48 &	 2.9E-3 &	 3.9E-12\\
2 &	18 June 2016 &	13.46 &	1473 &	1.76 &	 5.3E-3 &	 1.0E-11\\
3 &	19 June 2016 &	16.83 &	1546 &	1.69 &	 7.7E-3 &	 1.9E-11\\
4 &	20 June 2016 &	20.27 &	2003 &	1.98 &	 1.9E-3 &	 2.2E-13\\
5 &	09 June 2017 &	21.13 &	2015 &	1.48 &	 - 	  &		- \\
6 &	10 June 2017 &	17.68 &	1732 &	1.57 &	 6.2E-3 &	 9.2E-12\\
7 &	12 June 2017 &	10.97 &	1386 &	1.50 &	 3.4E-3 &	 4.7E-12\\
8 &	22 June 2017 &	26.34 &	2015 &	1.40 &	- 	  & 		- \\

\cline{1-7} 
    \multicolumn{7}{|c|}{Line of sight to the Crab includes a streamer} \\
        \cline{1-7}						
9 &	17 June 2016 &	10.20 &	845  &	2.44 &	- & -\\
10 &	17 June 2017 &	9.41  &	901  &	2.51 &	 - & - \\
11 &	18 June 2017 &	12.61 &	800  &	1.65 &	 - & - \\

\hline
\hline
\end{tabular}
\caption{The table describes the observational quantities and the derived plasma parameters in the solar wind. 
}
\label{tab:table-1}

\end{table}

\section{Observations through CMEs}\label{observations}
\begin{figure}[!ht]
	\centerline{\includegraphics[width=9.5cm]{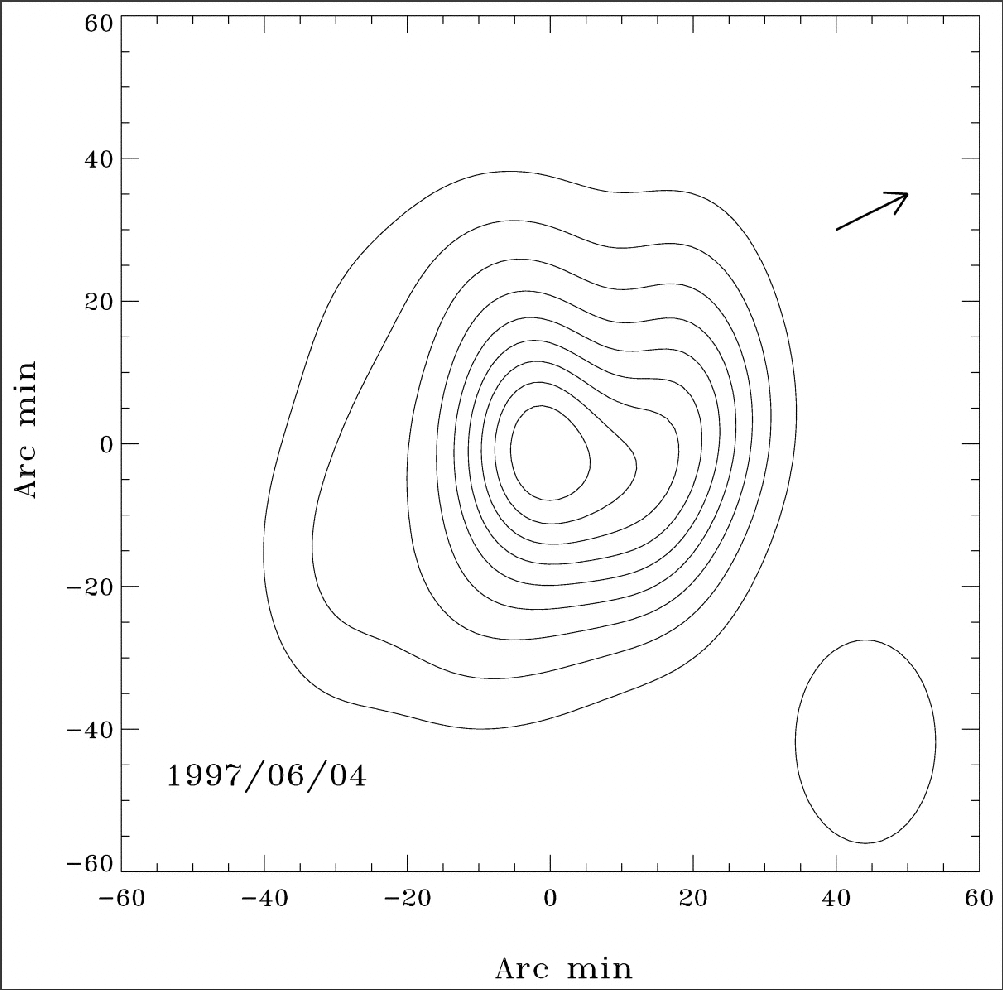}}
	\caption{Radio map of the Crab nebula obtained on 1997 June 4 with the CME is in the foreground at a heliocentric distance of $41 R_{\odot}$.}
	\label{fig:cmes}
\end{figure}

Low-frequency radio observations of the occultation of the Crab Nebula by the slow Coronal Mass Ejection event of 1997 June 2 were used to infer the physical properties of the CME at a heliocentric distance of approximately $41~R_{\odot}$. The observed angular broadening of the radio source is interpreted as being primarily caused by scattering due to plasma density inhomogeneities within the CME. Using such observations electron density of the CME was estimated and it was about 50 times greater than that of the surrounding coronal plasma at the same distance which is approximately $200~\mathrm{cm^{-3}}$; \citealt{Leb1998}. Further mass of the CME was inferred and mass is in good agreement with the values reported for some white-light CMEs at large heliocentric distances based on LASCO observations \citealt{Vou2000}, and supports the view that slow CMEs generally possess smaller masses \citep{Got1999}. These results demonstrate that such occultation measurements can provide an independent and complementary approach to investigating the connection between energetic phenomena in the solar atmosphere and the associated disturbances in the interplanetary medium. Figure \ref{fig:cmes} shows the scatter broadened image when there was a foreground CME at a distance of $41 R_{\odot}$.

\section{High frequency observations}\label{observations}
Very Large Array (VLA) observations in A configuration at wavelengths of 2, 3.5, 6, and 20~cm, shown the angular broadening of radio sources caused by the solar wind in the heliocentric distance range of 2--16~$R_{\odot}$. The observed angular broadening is anisotropic, with axial ratios ranging from 2 to 16. Larger axial ratios are preferentially detected at smaller solar elongations. Assuming that the anisotropy arises from scattering irregularities elongated along magnetic field lines, the distribution of position angles of the elliptically broadened images suggests that the magnetic field lines remain non-radial even at the largest heliocentric distances probed.Assuming a power-law spectrum for the electron density fluctuations, the fitted spectral indices lie in the range 2.8--3.4 for spatial scales between 2 and 35~km.

\begin{figure}[!ht]
	\centerline{\includegraphics[width=9.5cm]{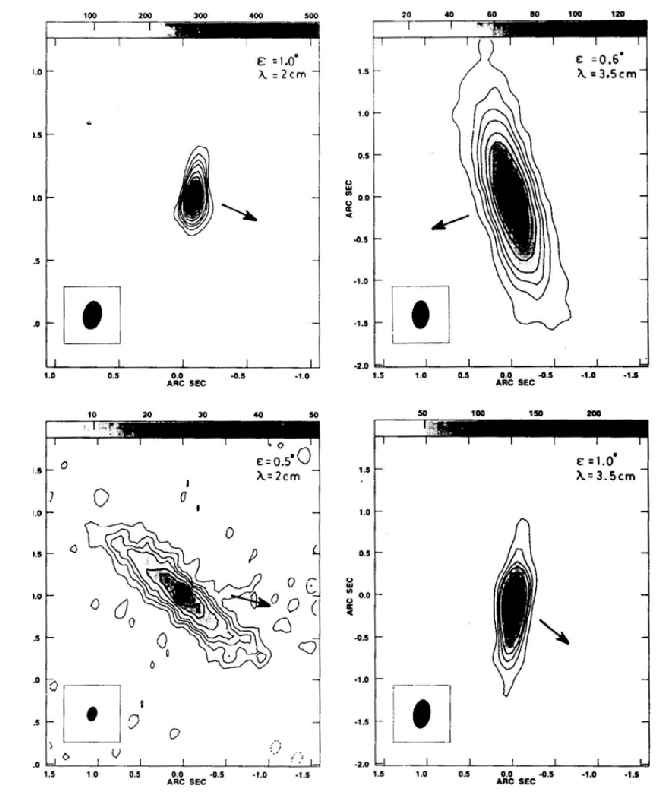}}
	\caption{Sample radio maps of different sources observed using VLA in A-configuration.}
	\label{fig:is}
\end{figure}

\section{Turbulence Regimes in Cometary Plasma Tails}\label{observations}

The random intensity fluctuation observed in signals received from a distant compact radio source, referred to as interplanetary scintillation (IPS, \cite{HSW64}), is due to diffraction and scattering caused by the turbulent and refracting solar wind plasma variations. IPS observations have allowed us to explore the solar wind throughout the inner heliosphere for several decades. Particularly, IPS observations at 327 MHz, facilitated by the action of the Fresnel filter, usually allow the tracking of solar wind electron density variations with scale sizes $\leq10^3$ km over heliocentric distances of 0.26 to 0.82 AU in the inner heliosphere \citep{RBA74}. IPS observations have been instrumental in measuring solar wind velocity \citep{Col96, MAn90, KKa90}, and also, the scintillation level measured using IPS acts as a perfect proxy for density variations in the inner heliospheric wind \citep{HTG85}. Thus, whenever interplanetary disturbances containing enhanced or depleted rms electron density fluctuations compared to the background solar wind cross the line‐of‐sight (LOS) to the observed source, they exhibit themselves as changes in the scintillation levels (m). In other words, whenever turbulence levels change in the solar wind, they will be reflected in IPS measurements as changes in m. 

IPS has been demonstrated to be a very efficient way to detect and characterize sub-arcsecond structures in radio sources. Assuming a Gaussian brightness distribution for the compact component of the radio source, the systematic variation of $m$ with $\epsilon$ in the weak scattering region can be used as an indicator of angular source size of scintillating radio source. An independent method of estimating angular diameter is based on fitting the observed IPS power spectra with solar wind parameters including the radial speeds, the strength of scintillation, inner and outer scales, power law index of density fluctuations along with the angular size of a scintillating component of the radio sources.

IPS is so sensitive to even small changes in density variations that it has been used to probe density fluctuations in tenuous cometary ion tails downstream of the comet nucleus. For example, enhanced scintillation levels were reported in comets Kohoutek \citep{ABR75}, Halley \citep{ABS86, SlM87, JaA92}, Wilson \citep{SlB90}, Austin \citep{JaA91}, Schwassmann-Wachmann 3-B \citep{RoM07}, ISON (C/2012 S1) \citep{IjA15}, and Neowise (C/2020 F3) \citep{FaF22}. Suppose density fluctuations similar to that in the solar wind plasma also exist in cometary plasma tails. In that case, the occultations of a compact radio source by the comet's plasma tail would produce enhanced scintillation similar to the IPS. However, comet observations with no enhanced scintillation level were also reported for comet Halley \citep{AMV87, HDu87}. This negative result was due to the fact that the cometary plasma tails get deflected from the true solar radial direction and lie along the resultant of the solar wind velocity vector and the comet velocity vector. This deflection, known as the tail lag, can be of the order of 3 degrees and if not accounted for the occultation event can be entirely missed.
Later, it has been shown conclusively \citep{SlB90} that the cometary plasma tail contains a higher turbulence level than the normal solar wind and produces enhanced scintillation.  Recently, a detailed occultation observation of a compact radio source by the plasma tail of comet C/2020 F3 (Neowise) was carried out with the Low-Frequency Array (LOFAR) \citep{FaF22}. The compact radio source 3C196 was almost perpendicularly behind the tail, providing a unique profile that showed unequivocally enhanced scintillation. \cite{FaF22} suggested that the enhanced scintillation was due to the scattering through the comet's plasma tail. Enhanced scintillation in the direction of a quasar has been reported at 103 MHz for comet Halley (Alurkar et al., 1986, Janardhan et al 1992) and for comet Austin during cometary tail occultations of strong radio quasars. It is important to note here that enhanced scintillation from cometary tail plasma can be observed only when the solar wind contribution is very low. Such a situation arises at large solar elongations when the solar wind contribution to the scintillation is small. The solar elongation during the observation was more than 60$^{\circ}$ \citep{JaA91}, highlighting the importance of ensuring the proper occultation geometry of the radio sources. The power spectra obtained by the occultation observation of 3C196 \citep{FaF22} showed the dual contributions from the solar wind and the plasma tail, illustrating the necessity for simultaneous data from multiple observations of additional radio sources not occulted by the plasma tail to separate the contributions of solar wind and plasma tail. Thus, more such occultation observation of compact radio sources through the cometary plasma tail is required. 

Comet occultation observations are rare. However, recently such opportunity arised to observe occultations of compact radio sources when Comet C/2023 A3 (Tsuchinshan-ATLAS) C/2025 N1 (3I/ATLAS) approached close to the Sun during September-October 2024 and October 2025. We wished to use the uGMRT facility, an interferometer, to observe the compact radio sources through the comet's tail using interferometric and phased array modes. Plasma density inhomogeneities in the solar wind scatter radio waves from compact radio sources and produce IPS. The occultation of compact radio sources by the comet's plasma tail will provide an excellent opportunity to use this phenomenon to probe turbulent density structure and dynamics in cometary ion tails. The confusion surrounding the contributions to enhanced scintillation can be unambiguously identified for the comet tail and separated from contributions from the solar wind in which it sits. Enhanced scintillation identified in the case of occultation observation of 3C196 by the plasma tail of comet C/2020 F3 (Neowise) \citep{FaF22} was attributed to the strong turbulence along the tail boundary, which is due to the substantial velocity shear exists between the tail boundary and the surrounding solar wind. One of our objectives will be to verify the turbulence nature in the plasma tail and find the reason behind it. Using the interferometric mode data obtained with the uGMRT, we wish to produce spatial correlation images that would further confirm the tail as the source of the enhanced scintillation. 


Using the regular IPS observations or the target of opportunity observations such as comet observations can be used for the estimation of angular source size of the radio sources. The size of compact radio sources can be determined by the interferometric observations obtained using the uGMRT. The size of a compact radio source can also be determined by systematically studying the behavior of the scintillation index (m) with solar elongation ($\epsilon$). The scintillation index (m) depends upon various factors such as observing frequency, elongation, the brightness distribution of the source B($\theta, \phi$), and the conditions in the interplanetary medium (IPM). However, at any given solar elongation a compact source will scintillate more that an extended source. Thus, a plot of m Vs elongation will be unique for a given souce size. This fact can be used to determine the angular sizes of radio sources using long-term IPS observations.

However, the apparent angular size of a compact radio source may appear significantly larger than its intrinsic size due to scattering effects in the interstellar medium. The apparent size $\theta_A$ of a radio source can be thought of as composed of two components, one as the intrinsic source size $\theta_I$ and one additional component $\theta_S$, arising due to scattering effects in the interstellar medium. Now the apparent size can be expressed as,

\begin{equation}
	\theta_A^2(\nu) = \theta_I^2(\nu) + \theta_S^2(\nu)
\end{equation}

Since $\theta_I$ is the intrinsic size of the source, it will be independent of the frequency and then the above equation can be used to estimate the angular broadening. Now, if we measure $\theta_A$ at two different frequencies $\nu_1$ and $\nu_2$, then angular broadening can be estimated as follow, 

\[\theta_A^2(\nu_1)-\theta_A^2(\nu_2) = \theta_S^2(\nu_1) - \theta_S^2(\nu_2)\]
\[\theta_A^2(\nu_1)-\theta_A^2(\nu_2) = \theta_S^2(\nu_1)[1 - \frac{\theta_S^2(\nu_2)}{\theta_S^2(\nu_1)}]\]

\begin{equation}
	\theta_A^2(\nu_1)-\theta_A^2(\nu_2) \approx \theta_S^2(\nu_1)
\end{equation}

using the values of apparent source sizes at 103 MHz at $\nu_1$ and at 151.5 MHz $\nu_2$, the angular broadening was estimated at 103 MHz by \citep{janardhan1993angular}. They concluded that interstellar scattering at 103 MHz is 0.07 ± 0.01, which is consistent with expectations from earlier studies at this frequency.

In summary, meter wavelength observations of cometary ion tail occultations of compact, extragalactic radio sources on a number of comets like Kohoutek, Halley, Wilson, Austin, and Neowise have led to a number of new insights into the nature of cometary tail plasma, well downstream of the cometary nucleus. In addition, comet Halley was observed at two different observing frequencies and when it was both close to the sun (0.2 AU and observed at 327 MHz) and far from the sun (1 AU and observed at 103 MHz). Collectively these observations have established the following – 1) Enhanced scintillations due to cometary ion plasma can only be observed under specific conditions of occulting geometry viz. at large solar elongations or large distances from the sun, when the solar wind contribution to the scintillation is very low. 2) Due to the comets velocity, in its path around the sun, and the radially directed solar wind, the position angle of the comets tail axis can be deflected by as much as $5^o$ \citep{1966ApJS...13..125B} from the true radial direction. Accounting for this tail lag is important as the occultation can be entirely missed if this so called `tail-lag' is not considered while computing occultation times. 3) Finally, such observations, which happen very rarely, have shown the existence of two distinct turbulence regimes in cometary tail plasma. One a small scale turbulence of 10 km to 40 km near the tail axis and a large scale turbulence of 100 km to 300 km at the edge or transition region where the cometary tail plasma merges with the solar wind. Correspondingly the rms electron density fluctuations in the two cometary regimes varied from around 0.4 $cm^{-3}$ to 5.0 $cm^{-3}$ respectively, a change of about an order of magnitude between the tail axis and the edge of the tail.

\section{Role of the SKAO on Occultation Observation}

As noted earlier, the SKAO is expected to offer a substantial enhancement over the capabilities of existing instruments, as well as its precursor and pathfinder facilities. By providing the sensitivity and imaging capability necessary to observe a large number of ecliptic sources in the solar vicinity at any given time, the SKAO will enable a new class of heliospheric and space weather studies. These observations will complement existing information derived from interplanetary scintillation (IPS) and heliospheric imaging, both under quiescent conditions and during dynamic events such as streamers, interaction regions, and coronal mass ejections (CMEs).

Such measurements will offer an unprecedented opportunity to characterize the turbulent coronal and solar wind plasma, as well as the associated magnetic fields, thereby contributing significantly to our understanding of solar wind heating and acceleration. The solar wind serves as an excellent natural laboratory for investigating the properties of magnetohydrodynamic (MHD) turbulence. Using radio occultation observations, it will be possible to explore several key aspects, including:

\begin{itemize}
\item Investigating the compressibility of solar wind turbulence.
\item Quantifying the role of radio wave scattering in producing the reduced quiet-Sun brightness temperatures observed at low radio frequencies.
\item Examining how solar wind turbulence influences the origin, coronal propagation, and apparent angular sizes of solar radio burst sources. Disentangling propagation effects from intrinsic source characteristics will enable a deeper understanding of the underlying emission mechanisms.
\item Determining the range of inner (dissipation) scales present in the solar wind. In order to fit the theoretical GSF profile, we need short base line observations which fulfilled by the SKAO.
\item Magnetic fields and magnetic field topology of the solar corona and solar wind by knowing the orientation of the semi major axis of the scatter broadened sources. 
\item Anisotropic behaviors of the solar corona and solar wind. 
\item Understanding the influence of turbulence on the propagation of Solar Energetic Particles (SEPs) through the heliosphere.
\item Radio occultation observations through coronal mass ejections and streamers will enable us to understand the various properites of them.
\item Long-term observations will allow us to investigate the properties of turbulence in the solar corona and their dependence on heliocentric distance, both in equatorial and polar regions. In other words, we will be able to probe turbulence characteristics in both the slow and fast solar wind. 
\item Turbulence levels in the cometary tails is another unique observational goals of SKAO. 
\item Density fluctuations inferred from the solar radio bursts observations in the solar corona can be used to derive the plasma heating rates \citep{Kon2025}.
\item High frequency angular broadening of SKAO will enable us to understand the innner scales close to the solar disk. 
\item Performing detailed comparisons of turbulence parameters derived from independent remote-sensing techniques such as radio occultation with in-situ measurements from missions like Parker Solar Probe, Solar Orbiter, and future space-based observatories.
\end{itemize}

The Square Kilometre Array Observatory will provide a transformative, unified view of solar wind turbulence and radio-wave propagation by combining ultra-high sensitivity, dense baseline coverage (including short baselines), and wideband imaging spectroscopy. It will enable precise quantification of the compressibility and anisotropy of turbulence, while simultaneously constraining density fluctuation spectra, inner (dissipation) scales, and their radial evolution through scattering, scintillation, and angular broadening measurements. By disentangling propagation effects from intrinsic emission, SKAO observations of solar radio bursts will refine our understanding of particle acceleration and emission mechanisms, and allow accurate inference of plasma heating rates. High-fidelity measurements of radio-wave scattering will resolve the long-standing issue of reduced quiet-Sun brightness temperatures at low frequencies. By knowing the semi-major axis of the solar corona, SKAO will be able to map the magnetic field strength and topology of the corona and solar wind. Radio occultation observations via CMEs, streamers, and even cometary tails will probe their internal structure and turbulence levels. Furthermore, by linking turbulence properties to the transport of Solar Energetic Particles (SEPs), SKAO will bridge plasma turbulence and space weather physics. Crucially, these remote-sensing diagnostics, when combined with in-situ measurements from missions such as the Parker Solar Probe and Solar Orbiter, will enable a self-consistent, multi-scale understanding of heliospheric plasma, marking a major step forward in solar and space physics.

\bibliographystyle{abbrvnat-maxbibnames4}
\bibliography{chapter} 

\end{document}